\documentclass[fleqn,usenatbib,useAMS]{mnras}

\usepackage{graphicx}
\usepackage{enumerate}
\usepackage{newtxtext,newtxmath}
\usepackage{ae,aecompl}
\usepackage[T1]{fontenc}
\usepackage{gensymb}
\usepackage{mathtools}
\usepackage{physics}
\usepackage{natbib}
\usepackage{hyperref}
\usepackage{color}
\usepackage{ulem}
\usepackage{soul}
\usepackage{url}
\usepackage{xspace}
\usepackage{multirow}
\usepackage{xcolor}
\usepackage{array}
\usepackage{makecell}
\usepackage[caption=false]{subfig}

\newcommand{\kms}{\,km\,s$^{-1}$} 

\newcommand{\changes}[1]{{#1}}

\title[WISDOM XXII.\ SMBH of NGC~383]{WISDOM Project -- XXII.\ A $5\%$ precision CO-dynamical supermassive black hole mass measurement in the galaxy NGC~383}

\author[H.\ Zhang et al.]{
Hengyue Zhang,$^{1}$\thanks{E-mail: hengyue.zhang@physics.ox.ac.uk}
Martin Bureau,$^{1}$\thanks{E-mail: martin.bureau@physics.ox.ac.uk}
Ilaria Ruffa,$^{2,3,6}$
Michele Cappellari,$^{1}$
Timothy A. Davis,$^{2}$
\newauthor{
Pandora Dominiak,$^{1}$
Jacob S. Elford,$^{2}$
Satoru Iguchi,$^{4,5}$
Federico Lelli,$^{6}$
Marc Sarzi, $^{7}$
Thomas G. Williams$^{1}$}
\\
$^{1}$Sub-department of Astrophysics, Department of Physics, University of Oxford, Denys Wilkinson Building, Keble Road, Oxford, OX1~3RH, UK \\
$^{2}$School of Physics \& Astronomy, Cardiff University, Queens Buildings, The Parade, Cardiff, CF24~3AA, UK \\
$^{3}$INAF - Istituto di Radioastronomia, via P.\ Gobetti 101, 40129 Bologna, Italy \\
$^{4}$Graduate Institute for Advanced Studies, SOKENDAI, Mitaka, Tokyo 181-8588, Japan\\
$^{5}$National Astronomical Observatory of Japan, National Institutes of Natural Sciences, Mitaka, Tokyo 181-8588, Japan\\
$^{6}$INAF, Arcetri Astrophysical Observatory, Largo Enrico Fermi 5, I-50125 Florence, Italy\\
$^{7}$Armagh Observatory and Planetarium, College Hill, Armagh BT61~9DG, UK
}

\pubyear{2024}

\begin{document}
\label{firstpage}
\pagerange{\pageref{firstpage}--\pageref{lastpage}}
\maketitle

\begin{abstract}
We present a measurement of the supermassive black hole (SMBH) mass of the nearby lenticular galaxy NGC~383, based on Atacama Large Millimeter/sub-millimeter Array (ALMA) observations of the $^{12}$CO(2-1) emission line with an angular resolution of $0\farcs050\times0\farcs024$ ($\approx16\times8$~pc$^2$). These observations spatially resolve the nuclear molecular gas disc down to $\approx41,300$ Schwarzschild radii and the SMBH sphere of influence by a factor of $\approx24$ radially, better than any other SMBH mass measurement using molecular gas to date. The high resolution enables us to probe material with a maximum circular velocity of $\approx1040$~\kms, even higher than those of the highest-resolution SMBH mass measurements using megamasers. We detect a clear Keplerian increase (from the outside in) of the line-of-sight rotation velocities, a slight offset between the gas disc kinematic (i.e.\ the position of the SMBH) and morphological (i.e.\ the centre of the molecular gas emission) centres, an asymmetry of the innermost rotation velocity peaks and evidence for a mild position angle warp and/or non-circular motions within the central $\approx0\farcs3$. By forward modelling the mass distribution and ALMA data cube, we infer a SMBH mass of \changes{$(3.58\pm0.19)\times10^9$~M$_\odot$} ($1\sigma$ confidence interval), more precise ($5\%$) but consistent within $\approx1.4\sigma$ with the previous measurement using lower-resolution molecular gas data. Our measurement emphasises the importance of high spatial resolution observations for precise SMBH mass determinations.
\end{abstract}



\section{Introduction}

Research in the past few decades has revealed tight correlations between supermassive black hole (SMBH) mass ($M_\mathrm{BH}$) and properties of the host galaxy such as stellar velocity dispersion, bulge mass and stellar mass \citep[e.g.][]{Ferrarese_2000, Gebhardt_2000, Beifiori_2012, Kormendy_2013}, thus providing strong evidence that SMBHs co-evolve with their host galaxies across cosmic time. It is now believed that feedback from active galactic nuclei (AGN) plays a crucial role driving and regulating this co-evolution, by changing the physical conditions of the interstellar medium (ISM) and/or removing it from the nuclear regions, thus quenching star formation (see e.g.\ \citealt{Alexander_2012}, \citealt{Morganti_2017} and \citealt{Harrison_2018} for reviews). Yet, the exact mechanisms underlying these processes remain unclear \citep[e.g.][]{Kormendy_2013, D'Onofrio_2021}. Moreover, increasing evidence suggests that SMBH -- galaxy correlations depend on the galaxy's morphological type and total stellar mass \citep[e.g.][]{McConnell_2013, Bosch_2016, Krajnovic_2018}, although the underlying causes are again uncertain. To constrain the mechanisms governing SMBH -- galaxy co-evolution and their dependence on galaxy type, obtaining many more precise and accurate SMBH mass measurements across a diverse range of galaxies is crucial.

Reliable SMBH mass measurements require spatially resolving and modelling the kinematics of matter within the SMBH's sphere of influence (SoI), where the SMBH gravitational influence dominates over that of other mass components (stars, gas, dust and dark matter). Common kinematic tracers include stars \citep[e.g.][]{Cappellari_2002b, Krajnovic_2009, Drehmer_2015}, ionised gas \citep[e.g.][]{Ferrarese_1996, Sarzi_2001, Walsh_2013} and megamasers (hereafter ``maser'' for short; e.g.\ \citealt{Herrnstein_2005, Kuo_2011, Gao_2017}). Measurements of SMBH masses using maser kinematics are currently considered the most precise and accurate \citep[e.g.][]{Kormendy_2013, Gao_2017}, as maser emission is typically observed using very long baseline interferometry (yielding angular resolutions much higher than those of other methods) and traces material very close to the SMBHs. However, the maser method is biased towards galaxies with $10^6\lesssim M_\mathrm{BH}\lesssim10^8$~M$_\odot$, as the required maser emission originates from a specific type of nuclear activity almost exclusively present in Seyfert~2 AGN of low-mass galaxies (see \citealt{Lo_2005} for a review). By contrast, measurements using stellar kinematics probe a wider range of SMBH masses but are biased towards relatively dust-free and non-disturbed objects with massive SMBHs. Most of these are early-type galaxies (ETGs; \citealt{Kormendy_2013}).

In recent years, substantial improvements in the sensitivity and angular resolution of sub-millimetre interferometers have enabled a new method to measure SMBH masses: probing the kinematics of molecular gas discs down to the SoI. This molecular gas method is well suited to a wide range of galaxy masses and nuclear activities, with $\approx35,000$ potential local targets \citep{Davis_2014}. It is however challenging to apply the technique to objects with prominent non-circular motions \citep[e.g.][]{Combes_2019} or central holes in their molecular gas discs \citep[e.g.][]{Kabasares_2022, Ruffa_2023}. The mm-Wave Interferometric Survey of Dark Object Masses (WISDOM) project, using high-resolution CO observations from primarily the Atacama Large Millimeter/sub-millimeter Array (ALMA), has so far provided accurate SMBH masses of eleven typical ETGs \citep{Davis_2013b, Onishi_2017, Davis_2017, Davis_2018, Smith_2019, North_2019, Smith_2021, Ruffa_2023, Dominiak_2024_submitted}, a dwarf ETG \citep{Davis_2020} and a peculiar luminous infrared galaxy with central spiral arms \citep{Lelli_2022}. Other groups have presented similar molecular-gas SMBH mass measurements of \changes{thirteen} additional ETGs \citep{Barth_2016, Boizelle_2019, Nagai_2019, Ruffa_2019b, Boizelle_2021, Cohn_2021, Kabasares_2022, Nguyen_2022, Cohn_2023, Cohn_2024, Dominiak_2024} and three late-type galaxies (LTGs, all barred spirals; \citealt{Onishi_2015, Nguyen_2020, Nguyen_2021}).

This paper presents a measurement of the SMBH mass of the ETG NGC~383 using ultra-high resolution ($\approx0\farcs034$) ALMA observations of the $^{12}$CO(2-1) emission line. A previous WISDOM study \citep{North_2019} inferred a SMBH mass of $(4.2\pm0.4)\times10^9$~M$_\odot$ using $\approx0\farcs13$ ALMA observations of the same line. According to \citet{Zhang_2024}, the physical scale probed by this prior measurement is only $\approx3$ times worse than that of the best SMBH measurement using masers \citep{Herrnstein_2005}, when evaluated in unit of the SMBH Schwarzschild radius ($R_\mathrm{Sch}\equiv2GM_\mathrm{BH}/c^2$, where $G$ is the gravitational constant and $c$ \changes{is} the speed of light). With $\approx4$ times better angular resolution, this new study aims to achieve the highest spatial resolution measurement (in the unit of $R_\mathrm{Sch}$) of all SMBH mass measurements to date, spatially resolving the central molecular gas disc with unprecedented detail to derive a much more precise SMBH mass. Section~\ref{sec:observation} introduces the target and presents our new ALMA observations and data reduction. We describe our molecular gas dynamical modelling technique to measure the SMBH mass in Section~\ref{sec:modelling} and discuss the sources of uncertainties and the importance of high-resolution observations in Section~\ref{sec:discussion}. We conclude briefly in Section~\ref{sec:conclusion}.


\section{ALMA Observations}
\label{sec:observation}

\subsection{Target: NGC~383}

NGC~383 is an unbarred dusty lenticular galaxy \citep{deVaucouleurs_1991} and the brightest galaxy of the NGC~383 group of galaxies, consisting of $11$ members \citep{Sakai_1994}. It is also the host galaxy of the radio source 3C~031, an AGN with relatively low-power radio jets showing a characteristic Fanaroff-Riley type I morphology (\citealp{Fanaroff_1974}) on kpc scales, with optical spectral properties typical of low-excitation radio galaxies (LERGs; see e.g.\ \citealt{MacDonald_1968, Bridle_1984, Laing_2002, vanVelzen_2012}). The combination of these characteristics makes 3C~031 the prototype of this class of AGN (see also \citealp{Ruffa_2020}).
We adopt a distance $D=66.6\pm9.9$~Mpc derived by \citet{Freedman_2001} using the Tully-Fisher relation \citep{Tully_1977}. At this distance, an angle of $1^{\prime\prime}$ corresponds to a spatial extent of $\approx 323$~pc.

The SMBH of NGC~383 has a large SoI. Adopting $M_\mathrm{BH}=(4.2\pm0.4)\times10^9$~M$_\odot$ from \citet{North_2019} and an effective (half-light) stellar velocity dispersion $\sigma_\mathrm{e}=239\pm16$~\kms\ from \citet{Bosch_2016}, we expect the radius of the SoI to be $R_\mathrm{SoI}\equiv GM_\mathrm{BH}/\sigma_\mathrm{e}^2=320$~pc or $\approx1\farcs0$. Our $0\farcs050\times0\farcs024$ (geometric average $\approx0\farcs034$) angular resolution ALMA observations presented in the next sub-section spatially resolve this SoI by a factor of $\approx29$ in radius ($\approx24$ for our updated $M_\mathrm{BH}$ in Section~\ref{subsec:results}), and should thus enable a precise SMBH mass determination.

The total molecular gas mass of NGC~383 is $(1.49\pm0.19)\times10^9$~M$_\odot$ \citep{OFlaquer_2010}, derived from single-dish observations of the $^{12}$CO(1-0) and the $^{12}$CO(2-1) lines with the Institut de Radioastronomie Millim{\'e}trique (IRAM) 30-m telescope. \citet{North_2019} however showed that the molecular gas mass enclosed in the nuclear region is negligible compared to the enclosed SMBH and stellar masses.


\subsection{Observations and data reduction}

We observed the $^{12}$CO(2-1) line of NGC~383 with ALMA using configuration C43-10 \changes{from} 2021 September 5 \changes{to 2021 September 9}, as part of project 2019.1.00582.S (PI: M.\ Bureau). The observations consist of three tracks with a total on-source integration time of $3.9$~h. The shortest and the longest baselines are $122$ and $16,196$~m, respectively, providing a maximum recoverable scale (MRS) of $0\farcs5$ ($0.16$~kpc). Because the MRS is smaller than the expected $R_\mathrm{SoI}$, we combine those data with two prior intermediate-resolution observing tracks to improve the $uv$-plane coverage and fully recover the SoI. One track (project 2015.1.00419.S, PI: T.\ Davis) was observed on 2016 June 21, using ALMA configuration C36-4 (baselines of $15$ to $704$~m). The other track (project 2016.1.00437.S, PI: T.\ Davis) was observed on 2017 August 16, using ALMA configuration C40-7 (baselines of $21$ to $3637$~m). These two lower-resolution tracks were used in \citet{North_2019} to derive the previous SMBH mass.  
The MRS of the combined observing tracks is $\approx2\farcs7$ ($\approx0.87$~kpc), approximately the size of the molecular gas disc as estimated by \cite{North_2019}.
\changes{The details of all observing tracks used in this study are summarised in Table~\ref{tab:Observing Tracks}.}

\begin{table*}
	\centering
	\caption{\changes{Properties of ALMA observing tracks.}}
	\label{tab:Observing Tracks}
	\begin{tabular}{lllllccl}
		\hline
		Project code & Track & Date & Config. & Baseline range & ToS & MRS & Calibration\\
            & & & & & (s) & (arcsec, kpc) & \\
            (1) & (2) & (3) & (4) & (5) & (6) & (7) & (8)\\ 
		\hline
		2015.1.00419.S & {\fontfamily{cmtt}\selectfont uid\_A002\_Xb499c3\_X377d} & 2016-06-21 & C36-4 & \phantom{0}15~m -- \phantom{0}0.7~km & \phantom{0}121 & 7.9, 2.6 & Pipeline, \textsc{CASA} 4.5.3 \\
        2016.1.00437.S & {\fontfamily{cmtt}\selectfont uid\_A002\_Xc36f2a\_X46c} & 2017-08-16 & C40-7 & \phantom{0}21~m -- \phantom{0}3.6~km & 1669 & 1.7, 0.5 & Pipeline, \textsc{CASA} 4.7.2 \\ 
        2019.1.00582.S & {\fontfamily{cmtt}\selectfont uid\_A002\_Xf031e1\_X16fe} & 2021-09-05 & C43-10 & 122~m -- 16.2~km & 4940 & 0.5, 0.2 & Excluded manually\\ 
        & {\fontfamily{cmtt}\selectfont uid\_A002\_Xf031e1\_X1dfe} & 2021-09-05 & C43-10 & 122~m -- 16.2~km & 4757 & 0.5, 0.2 & \makecell[tl]{Pipeline \& manual phase-only \\ self-calibration, \textsc{CASA} 6.2.1.7} \\ 
        & {\fontfamily{cmtt}\selectfont uid\_A002\_Xf06573\_X1594} & 2021-09-09 & C43-10 & 122~m -- 16.2~km & 4494 & 0.5, 0.2 & \makecell[tl]{Pipeline \& manual phase-only \\ self-calibration, \textsc{CASA} 6.2.1.7} \\ \hline
	\end{tabular} \\
        {\changes{{\sl Notes.} Columns: (1) Project code. (2) Track ID. (3) Observation date (year-month-day). (4) ALMA array. (5) Minimum and maximum baseline length. (6) Total on-source time (ToS). (7) Maximum recoverable scale (MRS), i.e.\ the largest angular scale that can be recovered with the given array. (8) Calibration method.}}
\end{table*}

\changes{Each of the five (three new and two previous) observing tracks has four spectral windows (SPWs).} A high-resolution SPW with a bandwidth of $1.875$~GHz ($\approx2500$~\kms) was centred on the redshifted $^{12}$CO(2-1) line frequency (\changes{$\approx227$}~GHz) in ALMA band~6. The new observations used $1920$ channels of $976.6$~kHz ($\approx1.3$~\kms), whereas the old observations used $3840$ channels of $488.3$~kHz ($\approx0.65$~\kms). Three additional SPWs, each of $2$~GHz with $128$ channels of $15.63$~MHz ($\approx20.3$~\kms), were \changes{employed} to detect continuum emission. 
\changes{The properties of the spectral windows of all adopted observing tracks are listed in Table \ref{tab:Spectral windows}.}
For all observing tracks, a standard calibration strategy was adopted, using a single bright quasar as both flux and bandpass calibrator and another one (or two) as phase calibrator. The flux and bandpass calibrator of the new high-resolution tracks presented here was J2253+1608. The phase calibrators were J0057+3021 and J0112+3208.

\begin{table*}
	\centering
	\caption{\changes{Properties of spectral windows of adopted observing tracks.}}
	\label{tab:Spectral windows}
	\begin{tabular}{llcccccc}
		\hline
		Project code & Track & SPW & Bandwidth & $N_\mathrm{channel}$ & Channel width & Central frequency & Flagged \\
            & & & (GHz, \kms) & & (MHz, \kms) & (GHz) & \\
            (1) & (2) & (3) & (4) & (5) & (6) & (7) & (8)\\ 
		\hline
		2015.1.00419.S & {\fontfamily{cmtt}\selectfont uid\_A002\_Xb499c3\_X377d} & 0 & 2\phantom{.00}, 2600 & \phantom{0}128 & 15.63, 20.3 & 228.6 & No \\ 
  	 & & 1 & 2\phantom{.00}, 2600 & \phantom{0}128 & 15.63, 20.3 & 241.8 & No \\ 
          & & 2 & 2\phantom{.00}, 2600 & \phantom{0}128 & 15.63, 20.3 & 243.8 & No \\ 
          & & 3 & 1.88, 2500 & 3840 & \phantom{0}0.49, \phantom{0}0.6 & 226.6 & No \\
        2016.1.00437.S & {\fontfamily{cmtt}\selectfont uid\_A002\_Xc36f2a\_X46c} & 0 & 2\phantom{.00}, 2600 & \phantom{0}128 & 15.63, 20.3 & 228.7 & No \\ 
  	 & & 1 & 2\phantom{.00}, 2600 & \phantom{0}128 & 15.63, 20.3 & 242.0 & No \\ 
          & & 2 & 2\phantom{.00}, 2600 & \phantom{0}128 & 15.63, 20.3 & 243.9 & No \\ 
          & & 3 & 1.88, 2500 & 3840 & \phantom{0}0.49, \phantom{0}0.6 & 226.8 & No \\
        2019.1.00582.S & {\fontfamily{cmtt}\selectfont uid\_A002\_Xf031e1\_X1dfe} & 0 & 2\phantom{.00}, 2600 & \phantom{0}128 & 15.63, 20.3 & 228.8 & Yes \\ 
  	 & & 1 & 2\phantom{.00}, 2600 & \phantom{0}128 & 15.63, 20.3 & 240.6 & Yes \\ 
          & & 2 & 2\phantom{.00}, 2600 & \phantom{0}128 & 15.63, 20.3 & 242.5 & Yes \\ 
          & & 3 & 1.88, 2500 & 1920 & \phantom{0}0.98, \phantom{0}1.3 & 226.8 & No \\
         & {\fontfamily{cmtt}\selectfont uid\_A002\_Xf06573\_X1594} & 0 & 2\phantom{.00}, 2600 & \phantom{0}128 & 15.63, 20.3 & 228.8 & No \\ 
  	 & & 1 & 2\phantom{.00}, 2600 & \phantom{0}128 & 15.63, 20.3 & 240.6 & No \\ 
          & & 2 & 2\phantom{.00}, 2600 & \phantom{0}128 & 15.63, 20.3 & 242.5 & No \\ 
          & & 3 & 1.88, 2500 & 1920 & \phantom{0}0.98, \phantom{0}1.3 & 226.8 & No \\ \hline
	\end{tabular} \\
        {\changes{{\sl Notes.} Columns: (1) Project code. (2) Track ID. (3) SPW ID. (4) Total bandwidth. (5) Number of channels. (6) Channel width. (7) Central frequency. (8) Whether the SPW is flagged manually.}}
\end{table*}

All \changes{the new high-resolution observing tracks} were calibrated with the \textsc{Common Astronomy Software Application} (\textsc{CASA}) package\footnote{\url{https://casa.nrao.edu/}} \citep{McMullin_2007}, version~6.2.1.7, using the standard pipeline provided by the ALMA Science Archive\footnote{\url{https://almascience.eso.org/aq/}}. While carefully inspecting the calibrated visibilities of the three tracks using the \textsc{CASA} task {\tt plotms}, prominent residual noise spikes were identified in some edge channels of the three continuum SPWs, which were flagged manually using the \textsc{CASA} task {\tt flagdata}. The diagnostic weblog generated by the ALMA pipeline for the same tracks also revealed that the weather conditions at the ALMA site were very poor during the observations. Combined with the very extended configuration of the array, this resulted in substantial residual phase noise (inducing significant amplitude decorrelation) after the standard calibration. A continuum image of each new track was thus created, revealing that only the first track was severely affected by the adverse weather conditions (showing substantial signal decorrelation and image artefacts). We therefore excluded this track from the rest of the analysis and combined the remaining calibrated (two new and two previous) tracks from all three ALMA projects using the \textsc{CASA} task {\tt concat}.

The analysis reported in the following was carried out using this combined dataset only, which was imaged using \textsc{CASA} version~6.5.2. All the data products were created with a H{\"o}gbom deconvolver \citep{Hogbom_1974}, a pixel size of $0\farcs01$ and an image size of $960\times960$ pixels, to reduce file sizes while sampling the synthesised beams properly.


\subsection{Continuum emission}

The continuum SPWs and the line-free channels of the line SPWs of the combined dataset were used to produce the continuum image. Various experiments with data reduction revealed that self-calibration\footnote{\url{https://casaguides.nrao.edu/index.php/First_Look_at_Self_Calibration_CASA_6.5.4}} could improve the quality of the data products. We thus performed two runs of phase-only self-calibration \changes{using the \textsc{CASA} task {\tt gaincal}}, progressively decreasing the solution interval from $5$ to $1$~min, leading to an increase of the dynamic range (i.e.\,the ratio between the peak flux and the root-mean-square noise) of the continuum image by a factor of about $4$ (from $\approx1000$ to $\approx4000$). The final continuum image was created using the \texttt{tclean} task in multi-frequency synthesis mode \citep{Rau_11}, with no spectral dependence, and Briggs weighting with a robust parameter of $2$ (equivalent to natural weighting). The latter choice was made to maximise the sensitivity and possibly image emission from the base of the radio jets, that are known to dominate the continuum spectrum of LERGs like NGC~383 down to mm wavelengths \citep[see e.g.][]{Ruffa_2019a, Ruffa_2020}. Indeed, the galaxy's radio-sub-mm spectral energy distribution was analysed by \citet{North_2019}, who reported a best-fitting power-law index $\alpha=-0.66\pm0.03$, consistent with synchrotron radiation from AGN radio jets \citep[see e.g.][]{Laing_2002,Ruffa_2020}.

\begin{figure}
    \centering
    \includegraphics[width=\linewidth]{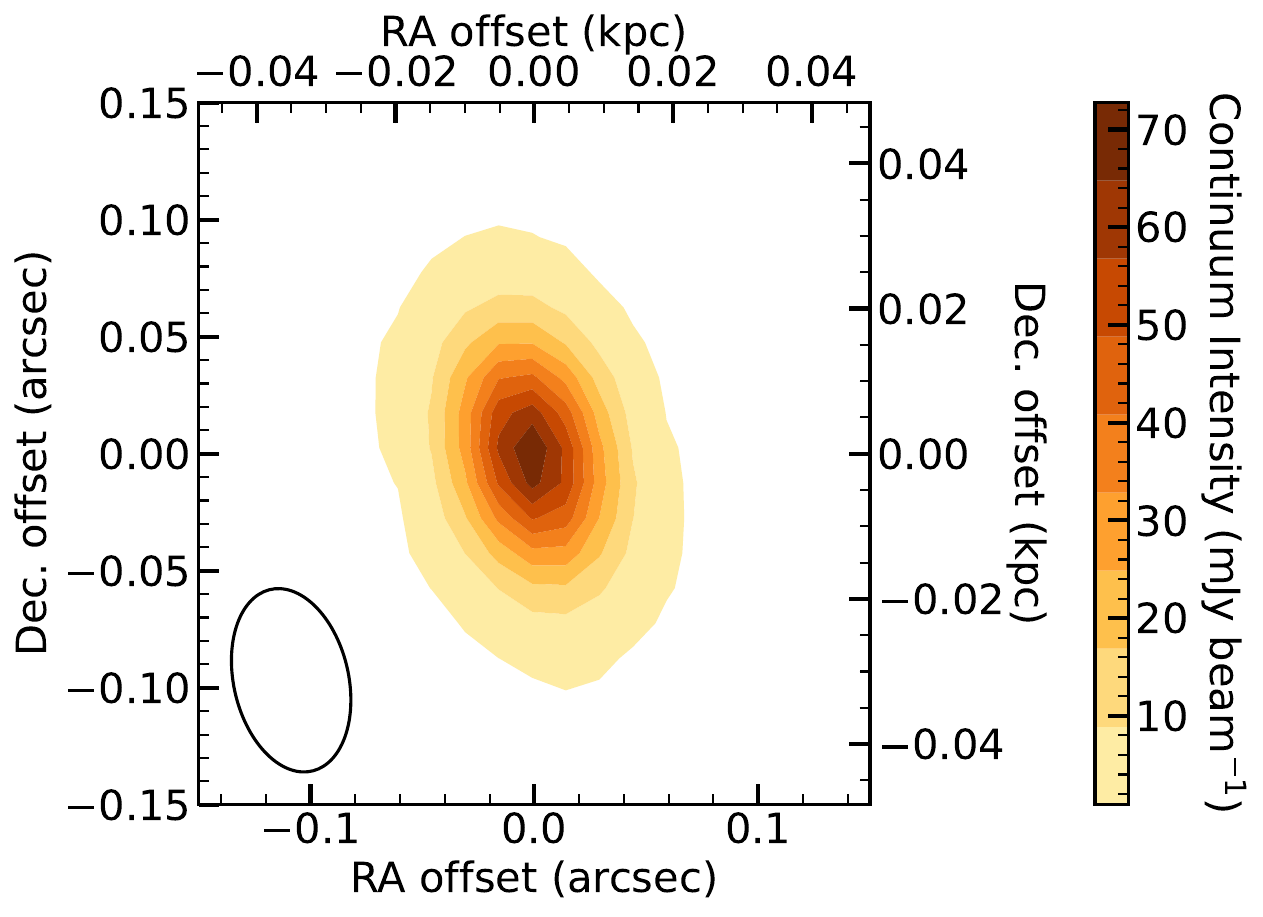}
    \caption{Central region (\changes{$0\farcs3\times0\farcs3$}) of the NGC~383 $1.3$~mm continuum image, showing the only (point) source detected. Contour levels are equally spaced between the peak intensity of $72.7\pm0.6$~mJy~beam$^{-1}$ and $50$ times the RMS noise. The synthesised beam ($0\farcs080\times0\farcs051$) is shown in the bottom-left corner as a black open ellipse.
    }
    \label{fig:continuum}
\end{figure}

The continuum image obtained has a root-mean-square (RMS) noise of $18$~$\mu$Jy~beam$^{-1}$ and a synthesised beam full-width at half-maximum (FWHM) of $0\farcs080\times0\farcs051$ ($\approx26\times16$~pc$^2$) at a position angle (PA) of $14\fdg3$. As shown in Figure~\ref{fig:continuum}, the continuum image reveals only one source near the galaxy's kinematic centre (best-fitting SMBH position; see Section~\ref{subsec:results}), with an integrated flux density of $71.0\pm0.6$~mJy, consistent (within the usual $\approx10\%$ ALMA flux calibration uncertainty) with that reported by \citet{North_2019}. The deconvolved size of the source, derived by fitting a two-dimensional (2D) Gaussian with the \textsc{CASA} task \texttt{imfit}, is consistent with it being spatially unresolved (i.e.\ a point source). \changes{The source remains unresolved if we image the continuum with a different robust parameter (e.g.\ $0.5$).}

\subsection{Line emission}
\label{subsec:line_em}

\subsubsection{Imaging}

After applying the continuum self-calibration to the line SPWs, the CO line emission was isolated in the $uv$-plane using the \textsc{CASA} task {\tt uvcontsub}, which forms a continuum model from linear fits to line-free channels in frequency and then subtracts this model from the visibilities. We then created the CO data cube using the {\tt tclean} task with Briggs weighting and a robust parameter of $0.5$, to maximise the angular resolution while achieving sufficient sensitivity. We adopted a channel width of $20$~km~s$^{-1}$, larger than that of \citeauthor{North_2019} (\citeyear{North_2019}; $10$~\kms) but necessary as the combined data cube used here generally has lower signal-to-noise ratios ($S/N$). The channel velocities were computed in the rest frame, a velocity of zero corresponding to the rest frequency of the $^{12}$CO(2-1) line (i.e.\ $230.538$~GHz). 
The continuum-subtracted dirty cube was cleaned in regions of line emission (identified interactively) to a conservative threshold of about $1.5$ times the RMS noise ($\sigma_\mathrm{RMS}$; measured from line-free channels).
The final, self-calibrated and cleaned $^{12}$CO(2-1) data cube has $\sigma_\mathrm{RMS}=0.14$~mJy~beam$^{-1}$ and a synthesised beam FWHM of $0\farcs050\times0\farcs024$ ($\approx16\times8$~pc$^2$) with a PA of $22\fdg5$, well resolving the SoI. The data cube dynamic range (peak $S/N$) is $\approx9$.


\subsubsection{Moment maps}

We visualise the final data cube by creating zeroth-moment (integrated-intensity) and first-moment (intensity-weighted mean line-of-sight velocity) maps using a masked moment technique \citep[e.g.][]{Dame_2011}. The mask is generated by taking a copy of the original data cube and smoothing that copy first spatially with a uniform filter and then spectrally with a Hanning window. Due to the small synthesised beam of our data cube, the $S/N$ ($\approx3$ per synthesised beam per $20$~\kms channel) are generally not as high as those of the \citeauthor{North_2019}'s (\citeyear{North_2019}) data cube. We thus chose the size of the filters ($3$ times the synthesised beam FWHM spatially and $11$ channels spectrally) and the clipping threshold ($0.5$~$\sigma_\mathrm{RMS}$ of the unsmoothed data cube or $\approx6.5$~$\sigma_\mathrm{RMS}$ of the smoothed cube) to optimise the trade-off between flux recovery and noise reduction. As a result, we smoothed over more pixels to suppress the noise in the moment maps and reduced the clipping threshold until most of the flux was recovered. All pixels in the smoothed cube above the threshold were then selected, and the moment maps were created from the unsmoothed cube using only those pixels. Irrespective of this procedure, we stress that the modelling described in Section~\ref{sec:modelling} uses the final unmasked data cube, not the moment maps.

\begin{figure*}
\captionsetup[subfigure]{labelformat=empty}
    \centering
    \subfloat[]{
    \includegraphics[width=0.478\linewidth]{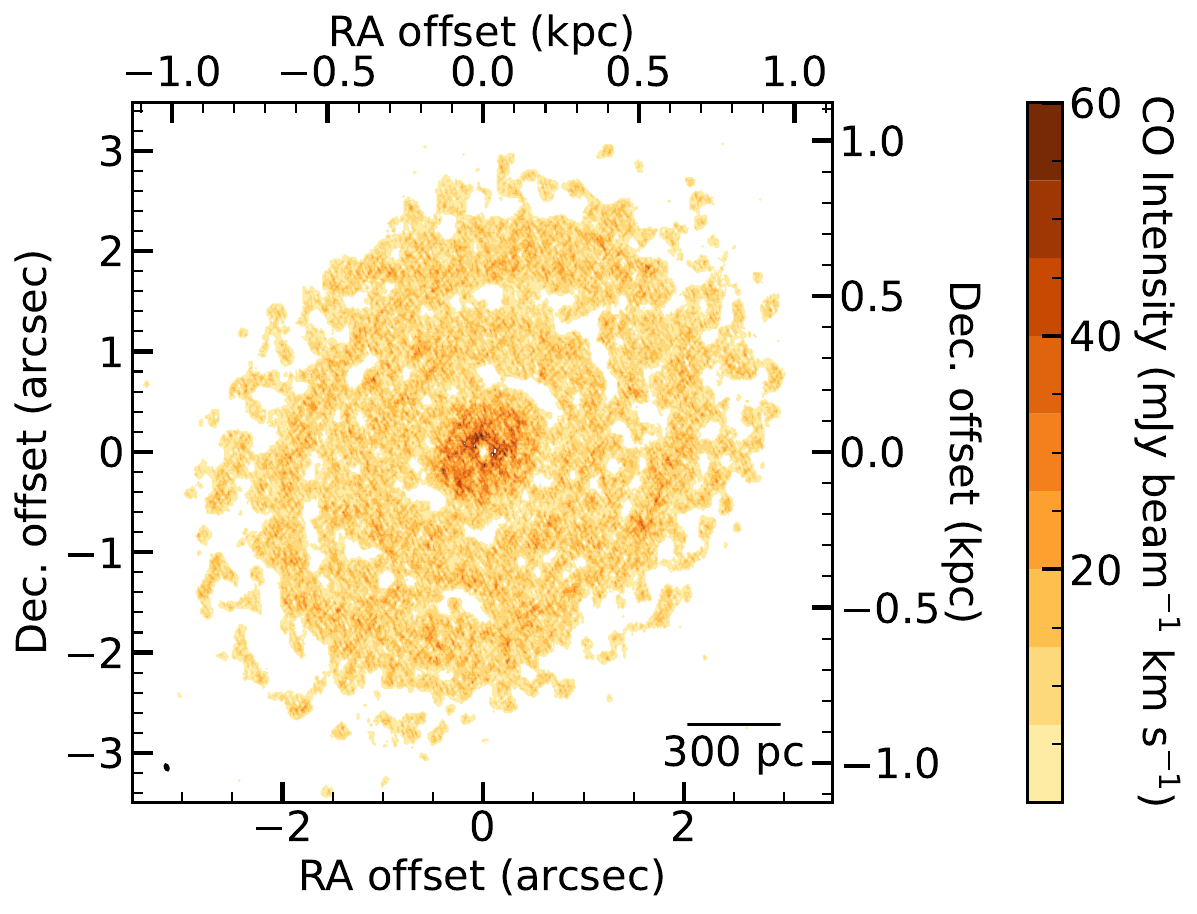}
    }
    \subfloat[]{
    \includegraphics[width=0.502\linewidth]{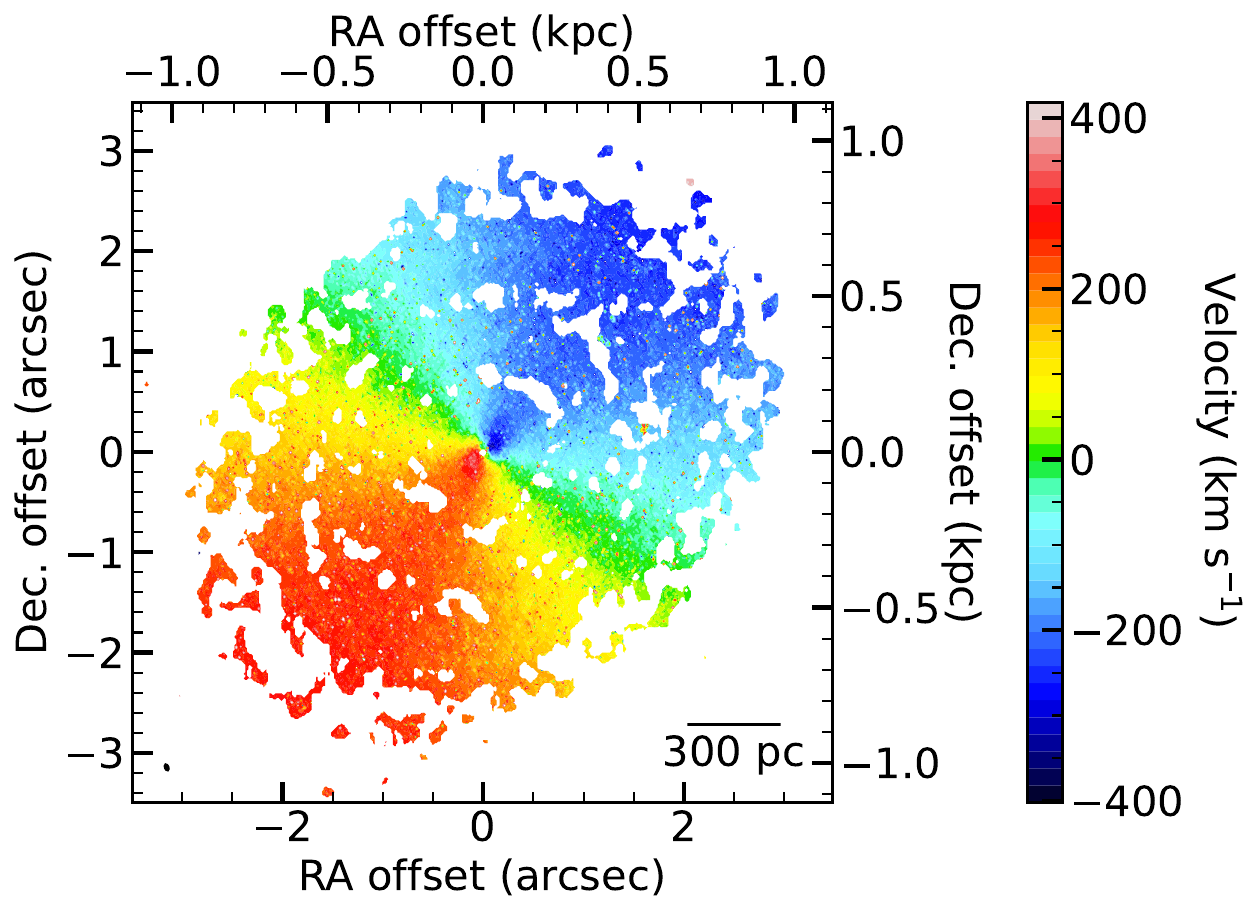}
    }
    \caption{Zeroth-moment (integrated-intensity; left panel) and first-moment (intensity-weighted mean line-of-sight velocity; right panel) maps of NGC~383 created from our ALMA $^{12}$CO(2-1) data cube. The synthesised beam ($0\farcs050\times0\farcs024$) is shown as a black filled ellipse in the bottom-left corner of each panel, while a $300$~pc scale bar is shown in the bottom-right corner of each panel. Positions are measured relative to the best-fitting kinematic centre; velocities relative to the best-fitting systemic velocity (see Section~\ref{subsec:results}).
    }
    \label{fig:moments}
\end{figure*}

Figure~\ref{fig:moments} shows the resulting moment maps. The molecular gas disc extends $\approx2$~kpc in radius along its major axis, and we recover the weak spiral structure seen in CO by \citet{North_2019}. The slight central dip in the zeroth-moment map is likely a physical decrease of the CO surface brightness within a radius of $\approx 0\farcs1$, rather than an artefact of the masking procedure as suggested by \citet{North_2019}, as the dip remains as the clipping threshold of the mask is decreased. \changes{Dips or holes are often present at the centres of molecular gas discs observed at high resolution \citep[e.g.][]{Davis_2018, Izumi_2020, Nguyen_2021, Ruffa_2023}. They are potentially due to the dissociation or excitation of CO caused by AGN activity \citep[e.g.][]{Izumi_2020} and/or the suppression of molecular cloud formation due to strong tidal forces \citep{Sarzi_2005}.} Nevertheless, the \changes{CO} central depression \changes{of NGC~383} does not impact the SMBH mass measurement, as there is sufficient emission within the fainter region to robustly trace the kinematics near the SMBH (see Section \ref{subsec:implications} for a zoomed-in view of the nuclear disc). The first-moment (mean velocity) map reveals a typical dynamically cold rotating disc, ideal for a SMBH mass measurement. The enhanced velocities at the centre of the velocity map clearly indicate the presence of a SMBH. In contrast to the previous observations of \citet{North_2019} suggesting an unwarped disc, our high-resolution observations reveal a slight twist of the isovelocity contours within the central $\approx0\farcs3$ in radius, indicative of a PA warp and/or non-circular motions. We present models of this feature in Section~\ref{subsec:non-circ}. Other features match those recovered at lower angular resolution by \citet{North_2019}.

\subsubsection{Total flux}

Figure~\ref{fig:spec} shows the integrated $^{12}$CO(2-1) spectrum of NGC~383, extracted from the central $6^{\prime\prime}\times6^{\prime\prime}$ region of our high-resolution data cube, thus covering all of the detected emission. The spectrum clearly shows the typical double-horned shape of a rotating disc, also observed by \citet{Lim_2000}, \citet{Okuda_2005} and \citet{North_2019}. The total $^{12}$CO(2-1) flux in the cleaned data cube is $128.9\pm1.3$~Jy~\kms. This is about 50\% larger than the total flux of $87.1$~Jy~\kms\ reported by \citet{North_2019}. The flux discrepancy becomes even larger if we perform another \texttt{tclean} run with a shallower cleaning threshold of $2$~$\sigma_{\rm RMS}$, the total flux in the resultant data cube then increasing to $161.0\pm1.3$~Jy~\kms. This flux discrepancy is due to a known issue \citep[e.g.][]{Jorsater_1995, Bureau_2002} when evaluating total fluxes of certain multi-configuration datasets. When the dirty beam has a highly irregular shape, its effective area can differ substantially from that of the (clean) Gaussian beam of equivalent angular resolution, causing the true total flux in the residual map of \texttt{tclean} (hereafter "residual flux" for short) to be lower than the measured residual flux by a factor
\begin{equation}
\epsilon=\frac{\mathrm{clean\,beam\,area}}{\mathrm{dirty\,beam\,area}}\,\,.
\end{equation}
Here, $\epsilon$ is thus the correction factor required to obtain the true residual flux. The true total flux $T$ is thus related to the total flux in the CLEAN components $C$ and the measured residual flux $R$ by 
\begin{equation}
T=C+\epsilon R\,\,,
\end{equation}
while the measured total flux $M$ is
\begin{equation}
M=C+R\,\,.
\end{equation}
For a data cube produced with a deeper CLEAN, the discrepancy between $M$ and $T$ is smaller, as less flux remains in the residual map. Thus $M$ should converge asymptotically to $T$ for a sufficiently deep cleaning threshold ($\ll1$~$\sigma_\mathrm{RMS}$). Such a deep CLEAN is however impractical, as it introduces numerous artefacts in the output data cube, so we instead compute $T$ and $\epsilon$ using the two different values of $C$ and $R$ from our two \texttt{tclean} runs, following the prescription of \cite{Jorsater_1995}:
\begin{align}
\epsilon&=\frac{C_2-C_1}{R_1-R_2} \\
T&=\frac{R_1C_2-C_1R_2}{R_1-R_2}\,\,.
\end{align}
Substituting $C_1=61.5$~Jy~\kms\  and {$R_1=67.4\pm1.3$~Jy~\kms\  from the deeper \texttt{tclean} run, and $C_2=51.4$~Jy~\kms\  and $R_2=109.5\pm1.3$~Jy~\kms\  from the shallower \texttt{tclean} run, we obtain $\epsilon=0.24\pm0.01$ and $T=77.6\pm0.9$~Jy~\kms. We verified this result by creating two other data cubes cleaning down to thresholds of $1$~$\sigma_\mathrm{RMS}$ and $0.7$~$\sigma_\mathrm{RMS}$, respectively. The $\epsilon$ and $T$ derived from those two additional data cubes agree well with those above.

We note that the total flux of each data cube used in this flux correction procedure is obtained by integrating over the channels of the integrated spectrum with $S/N\geq3$. Integrating over the full velocity range of each data cube yields a consistent total flux within the statistical uncertainties. Our computed total $^{12}$CO(2-1) flux is slightly smaller than that reported by \cite{North_2019}, but it is consistent with the single-dish $^{12}$CO(2-1) flux of $74.4\pm2.8$~Jy~\kms\ from \citet{OFlaquer_2010} considering ALMA's flux calibration uncertainty ($\approx10\%$). This implies that the fluxes reported by \cite{North_2019} might also have been slightly overestimated, for similar reasons. In any case, our SMBH mass measurement is insensitive to this flux rescaling, so we proceed with the conservatively cleaned data cube (with a threshold of $1.5$~$\sigma_\mathrm{RMS}$) to minimise the artefacts introduced by \texttt{tclean}.

\begin{figure}
    \centering
    \includegraphics[width=\linewidth]{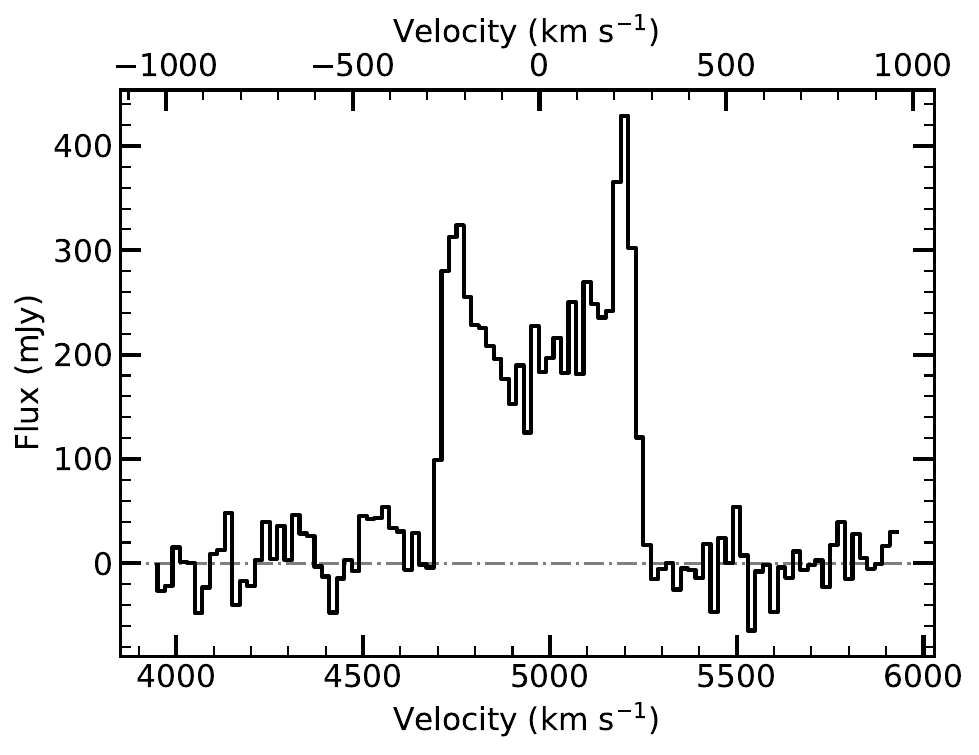}
    \caption{Integrated $^{12}$CO(2-1) spectrum of NGC~383, extracted from the central $6^{\prime\prime}\times6^{\prime\prime}$ region of our high-resolution data cube, covering all of the detected emission. Velocities are measured relative to the best-fitting systemic velocity along the top axis (see Section~\ref{subsec:results}). The dot-dashed line indicates the zero flux level. The spectrum shows the typical double-horned shape of a rotating disc.}
    \label{fig:spec}
\end{figure}


\subsubsection{PVD}

Figure~\ref{fig:pvd} shows the kinematic major-axis position-velocity diagram (PVD) of NGC~383, extracted along a PA of $142\degree$ (the best-fitting PA obtained in Section~\ref{subsec:results}) by summing the flux within a $15$-pixel wide pseudo slit. When creating PVDs, we adopt a masking procedure slightly different from that used to create the moment maps, to avoid masking out the central region: we use a spatial Gaussian filter of FWHM equal to that of the synthesised beam (rather than the larger uniform filter). We then select all pixels in the smoothed cube above $0.5$~$\sigma_\mathrm{RMS}$ of the unsmoothed data cube. Figure~\ref{fig:pvd} shows only the central $1\farcs5$ on either side of the kinematic centre, revealing a sharp Keplerian rise (from the outside in) of the line-of-sight velocities within the central $\approx0\farcs5$ in radius, the characteristic kinematic signature of a SMBH. As our high-resolution observations spatially resolve material $\approx4$ times closer to the SMBH than the previous observations of \citet{North_2019}, we detect velocities up to $\approx635$~\kms\ in the blueshifted half of the disc, $\approx1.8$ times larger than the highest velocity previously detected ($\approx350$~\kms).

A Keplerian circular velocity curve $v_\textrm{circ}(r)\propto r^{-1/2}$, where $v_\textrm{circ}(r)$ is the circular velocity at radius $r$, and the four times better angular (and thus spatial) resolution, suggest we should detect a maximum velocity of $\approx700$~\kms\ (twice the previous maximum velocity). Features up to $\approx700$~\kms\ are indeed detected if we lower the clipping threshold of the PVD, but those features are indistinguishable from noise due to the low $S/N$. Hence, we do not claim detection of such high velocities.

Noticeably, our high-resolution observations reveal a mild asymmetry in the detected velocity peaks at small radii; the redshifted component of the molecular gas disc only reaches $\approx535$~\kms. This asymmetry could arise from non-circular motions (e.g.\ gas inflows/outflows) or a deficiency of gas in the redshifted component of the nuclear disc (due to a specific gas morphology, e.g.\ a nuclear spiral). We discuss this further in Section~\ref{subsec:non-circ}. Another noticeable feature is that a small amount of central emission extends beyond the major-axis position of zero to the opposite side of the disc (i.e.\ into the "forbidden quadrants" of the PVD). This is consistent with the observed central twist of the isovelocity contours of the velocity map, again indicating a possible PA warp and/or non-circular motions. We attempt to reproduce this feature with 3D disc models in Section~\ref{subsec:non-circ}.

\begin{figure}
    \centering
    \includegraphics[width=\linewidth]{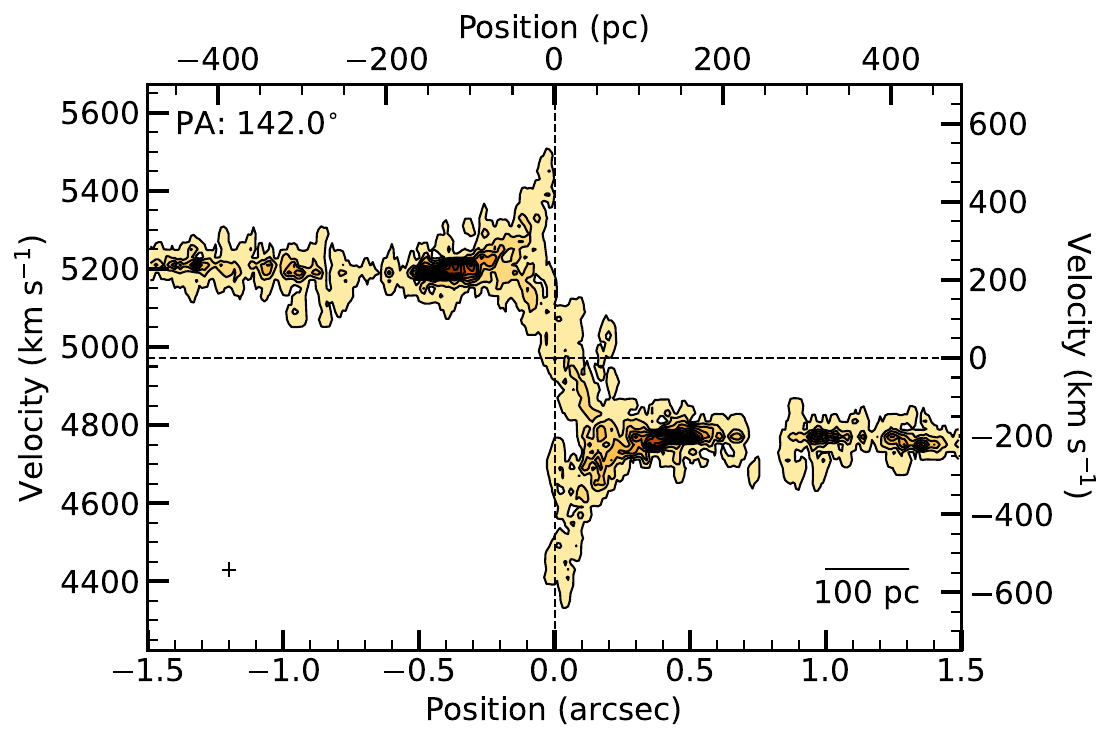}
    \caption{Major-axis position-velocity diagram of NGC~383, covering the central $1\farcs5$ on either side of the kinematic centre. The sharp increase (from the outside in) of the line-of-sight velocities within the central $\approx0\farcs5$ is a clear kinematic signature of a central SMBH. Positions are measured relative to the best-fitting kinematic centre; velocities relative to the best-fitting systemic velocity along the right axis (see Section~\ref{subsec:results}). The cross in the bottom-left corner shows the synthesised beam FWHM along the kinematic major axis and the channel width. Velocities up to $\approx635$~\kms\ are detected, $\approx1.8$ times larger than the highest velocity ($\approx350$~\kms) probed by the earlier observations of \citet{North_2019}.}
       \label{fig:pvd}
\end{figure}


\section{Dynamical Modelling}
\label{sec:modelling}

Our method to measure the SMBH mass of NGC~383 has been used and described extensively in previous WISDOM papers \citep[e.g.][]{Davis_2017, Smith_2019, North_2019, Ruffa_2023}, but as is often the case we need to make slight adjustments to these procedures for the specific galaxy being studied. We thus summarise the method and discuss the specifics of the NGC~383 modelling in this section.

We analyse the observed gas kinematics using the publicly available \textsc{Python} version of the \textsc{Kinematic Molecular Simulation} tool\footnote{\url{https://github.com/TimothyADavis/KinMS_fitter}\\
 \phantom{xx}\url{https://kinms.space/}} (\textsc{KinMS}; \citealt{Davis_2013a}). This takes an input model of the gas distribution and kinematics to create a mock data cube, taking into account beam smearing, spatial and velocity binning, and line-of-sight projection effects. The simulated data cube can then be compared directly to the observed data cube to infer the best-fitting parameters of the input model as well as their uncertainties, calculated via a Markov chain Monte Carlo (MCMC) $\chi^2$ minimisation routine. 

When computing the $\chi^2$, we rescale the uncertainties of the data cube by a factor of $(2N)^{0.25}$, where $N=404,722$ is the number of pixels with detected emission, defined as the pixels included in the mask of Section~\ref{subsec:line_em}. This approach attempts to account for the potential systematic uncertainties usually dominating large data sets, leading to more realistic fit uncertainties. The idea was proposed by \cite{vdBosch_2009} when using $\chi^2$ confidence levels and was later adapted by \cite{Mitzkus_2017} for Bayesian methods. We have since adopted this approach in various WISDOM papers \citep[e.g.][]{Smith_2019,North_2019}.
In addition, we assume a diagonal covariance matrix, ignoring the correlation between nearby pixels. Applying the full covariance matrix would be computationally intractable, and the impact would be much smaller than that of the rescaling described above (see \citealt{Davis_2017} for an extensive discussion of this issue).


\subsection{Mass model}
  
To generate a model of the gas kinematics, we assume that the gas particles are in circular motions (we explore the validity of this assumption in Section~\ref{subsec:non-circ}), with the circular velocity curve determined only by the mass distribution of the galaxy. When creating the mass model, we assume that the gravitational potential within the region of interest is dominated by the stars and the SMBH, ignoring contributions from other possible mass components (e.g.\ molecular gas, ionised gas, dust and dark matter). This assumption is reasonable as the molecular gas mass of NGC~383 is insignificant within the SMBH SoI \citep{North_2019}, and dark matter is usually negligible in the nuclear regions of massive galaxies \citep[e.g.][]{Cappellari_2013, Zhu_2024}.

\changes{We construct} our model of the stellar mass distribution \changes{using an approach similar to} that of \citet{North_2019}. The stellar light distribution is modelled by performing a multi-Gaussian expansion (MGE; \citealt{EMB_1994}), using the Python version of the \texttt{mge\_fit\_sectors\_regularized} procedure\footnote{Version 5.0 from \url{https://pypi.org/project/mgefit/}} of \citet{Cappellari_2002a} on a combined \textsl{Hubble Space Telescope} (\textsl{HST}) Near Infrared Camera and Multi-Object Spectrometer F160W-filter and Two Micron All Sky Survey \changes{(2MASS)} $H$-band image. We mask the \textsl{HST} image to exclude regions obscured by dust \changes{and mask the 2MASS image to remove pixels contaminated by a nearby star}, using the same mask\changes{s} as \cite{North_2019}. The combined masked image is thus fit by a sum of 2D Gaussians that are then analytically deprojected to a three-dimensional (3D) axisymmetric light distribution (given a free inclination $i$).
We list the parameters of each spatially-deconvolved Gaussian of the MGE model in Table~\ref{tab:MGE}.

\begin{table}
    \centering
    \caption{MGE parameterisation of the NGC~383 stellar light distribution.}
    \begin{tabular}{c@{\hskip 1cm}c@{\hskip 1cm}c}
      \hline
      $I_j$ & $\sigma_j$ & $q_j$ \\
      ($L_{\odot,\rm F160W}$~pc$^{-2}$) & (arcsec) & \\
      (1) & (2) & (3) \\
      \hline
      \changes{5789.27} & \changes{\phantom{0}0.114} & \changes{0.9\phantom{00}} \\
      \changes{2359.50} & \changes{\phantom{0}0.836} & \changes{0.95\phantom{0}} \\
      \changes{6889.76} & \changes{\phantom{0}1.07\phantom{0}} & \changes{0.95\phantom{0}} \\
      \changes{4379.08} & \changes{\phantom{0}2.34\phantom{0}} & \changes{0.95\phantom{0}} \\
      \changes{3535.85} & \changes{\phantom{0}4.61\phantom{0}} & \changes{0.903} \\
    \changes{\phantom{0}999.79} & \changes{12.5\phantom{00}} & \changes{0.95\phantom{0}} \\
      \hline
    \end{tabular}\\
    {{\sl Notes.} Parameters of the deconvolved Gaussian components. (1) Surface brightness. (2) Standard deviation (width). (3) Axial ratio.}
    \label{tab:MGE}
\end{table}

Our MGE model yields a light profile almost identical to that used by \cite{North_2019}, \changes{except for the innermost Gaussian component}. \changes{The MGE components published by \cite{North_2019} were deconvolved by the point-spread function (PSF) of the 2MASS image, mistakenly even in regions of the \textsl{HST} image, resulting in an unrealistically narrow innermost component.} By contrast, our listed MGE components have been deconvolved by the \changes{\textsl{HST}} PSF calculated using the \textsc{TinyTim} package\footnote{Version 7.5 from \url{https://github.com/spacetelescope/tinytim/releases/tag/7.5}} \citep{Krist_2011}. \changes{Nevertheless, the difference between the integrated masses of the previous and the new innermost Gaussian components is less than $20\%$ of the SMBH mass uncertainty derived in Section~\ref{subsec:results} and less than $7\%$ of the difference between the SMBH masses derived from the intermediate-resolution data and new high-resolution data. In the comparison of the statistical and systematic uncertainties of the two measurements presented in Section~\ref{subsec:uncertainty}, this small difference is negligible.} 

The circular velocity curve corresponding to the 3D stellar light distribution is computed using the \texttt{mge\_circular\_velocity} procedure of the \textsc{Jeans Anisotropic Modelling} (\textsc{JAM}) package\footnote{Version 7.2 from \url{https://pypi.org/project/jampy/}} of \citet{Cappellari_2008}. This procedure assumes a solar mass-to-light ratio ($M/L$), so we multiply the circular velocity generated \changes{with this procedure} by the square root of our (free) $M/L$. As in \citet{North_2019}, we adopt a linearly varying $M/L$:
\begin{equation}
    M/L(R)=M/L_\mathrm{inner}+(M/L_\mathrm{outer}-M/L_\mathrm{inner})\left(\frac{R}{3.5~\mathrm{arcsec}}\right)\,\,\,,
\end{equation}
where $R$ is the cylindrical radius, $M/L_\mathrm{inner}$ is the $M/L$ at $R=0$ and $M/L_\mathrm{outer}$ is the $M/L$ at $R=3\farcs5$ (approximately the outer edge of the gas disc). The inner and the outer $M/L$ are free parameters of our model, and the $M/L$ is constant (at $M/L_\mathrm{outer}$) at $R>3\farcs5$. 

Finally, we add a central point mass representing the SMBH and compute the circular velocity curve resulting from both the stars and the SMBH. \textsc{KinMS} then combines the circular velocity curve with a variable velocity dispersion to model the kinematics of the gas distribution.


\subsection{Gas distribution}

The mild spiral structure in the molecular gas disc of NGC~383 makes it inappropriate to assume a smooth axisymmetric gas distribution described by a simple parametric function (e.g.\ an exponential disc). Instead of attempting to construct a complicated function describing this 2D light/mass density distribution (that would necessarily have a large number of parameters), we simply adopt the observed but spatially-deconvolved molecular gas distribution as the input. Using the \textsc{SkySampler} tool\footnote{\url{https://github.com/Mark-D-Smith/KinMS-skySampler}} developed by \citet{Smith_2019}, we thus uniformly sample the (deconvolved) CLEAN components produced by the \textsc{CASA} task \texttt{tclean}, and generate $4.1\times10^6$ gas particles that exactly replicate the observed CO surface brightness distribution when convolved by the synthesised beam. The gas particles are then passed to \textsc{KinMS} with the position, systemic velocity, PA and inclination of the gas disc (assumed to be equal to those of the stars) as free parameters. Our procedure is identical to that used by \citet{North_2019} but uses the CLEAN components of our new combined observations rather than the old ones.


\subsection{Nuisance parameters}

Our dynamical modelling involves four nuisance parameters: the two coordinates of the kinematic centre (where the SMBH is) relative to the phase centre of the data cube, the systemic velocity of the gas disc and the integrated intensity of the CLEAN components (i.e.\ the scaling factor of the input gas distribution). We note that previous WISDOM papers assumed that the kinematic centre coincides with the gas disc's morphological centre (the centre of the molecular gas emission). This assumption was usually valid as the separation between the kinematic and the morphological centres was often much smaller than the synthesised beam size. However, in our high-resolution data cube, the two positions are more than one synthesised beam apart. We thus fit the two positions separately.

As we adopt the CLEAN components as the input gas distribution model, we can directly compute the position of the morphological centre relative to the data cube's phase centre, by calculating the intensity-weighted average position of the gas particles sampled from the CLEAN components. In this way, we confirm that this offset is less than $1\%$ of the synthesised beam FWHM, so from now on we assume that the morphological centre coincides with the data cube's phase centre. This leaves only the coordinates of the kinematic centre (relative to the phase centre) as free parameters, speeding up convergence.


\subsection{Prior distributions}

Although our data cube has a substantially higher angular and thus spatial resolution than the previous one from \citet{North_2019}, a MCMC fit with uninformative priors does not guarantee a more precise SMBH mass because of the lower $S/N$ of our data cube (typically $\approx3$ per synthesised beam per $20$~\kms\ channel). Indeed, the lower $S/N$ worsens the constraint on the inclination, which tightly correlates with the SMBH mass (through the deprojection of the velocities, i.e.\ $M_\mathrm{BH}\propto\sin^{-2}i$) and dominates the error budget (see Section~\ref{subsec:other_uncertainty}). Hence, we place informative priors on the inclination and the $M/L$ using constraints derived from fitting the previous intermediate-resolution data cube. 

Rather than directly adopting the results of \citet{North_2019} as priors, we re-analyse the intermediate-resolution data cube with our modified code that does not assume an overlap of the molecular gas kinematic and morphological centres. Table~\ref{tab:results} lists the best-fitting parameters and associated uncertainties resulting from fitting the intermediate-resolution data cube with our modified code. The best-fitting SMBH mass \changes{$\log\left(M_\mathrm{BH}/\mathrm{M}_\odot\right)=9.59\pm0.05$} ($1\sigma$ uncertainty here and throughout this paper), consistent within $1\sigma$ with the SMBH mass $\log\left(M_\mathrm{BH}/\mathrm{M}_\odot\right)=9.63\pm0.04$ of \citet{North_2019} (here and throughout this paper, when we evaluate the difference of two measurements, we adopt an uncertainty $\sigma=\sqrt{\sigma_1^2+\sigma_2^2}$, where $\sigma_1$ and $\sigma_2$ are the uncertainties of the first and the second measurement, respectively). The slight difference is statistically insignificant but could be the result of separating the kinematic and morphological centres \changes{and/or the improved MGE model}. The systemic velocity of $4977\pm1$~\kms\ is significantly higher than that of \citeauthor{North_2019} (\citeyear{North_2019}; $4925\pm1$~\kms), as we have corrected a previous inaccuracy in imaging the intermediate-resolution data: when converting from frequency to velocity, the previous imaging pipeline adopted a rounded rest frequency of $230.5$~GHz for the $^{12}$CO(2-1) line, instead of the more accurate $230.538$~GHz, causing a (non-physical) velocity offset of $\approx49$~\kms. The best-fitting systemic velocity of \citet{North_2019} agrees with ours after accounting for this inaccuracy.

We then adopt the posterior distributions of the inclination and the inner and outer $M/L$ as priors when fitting the high-resolution data cube. We use flat priors for all other parameters except the SMBH mass, for which we use a flat logarithmic prior. The ranges of the flat priors are listed in the second column of Table~\ref{tab:results}.

\begin{table*}
    \centering
    \caption{Best-fitting parameters and associated uncertainties, from the fits to the intermediate-resolution and high-resolution data cubes.}
    \begin{tabular}{lc@{\hskip 0.45cm}c@{\hskip 0.45cm}c@{\hskip 0.45cm}c@{\hskip 0.45cm}c@{\hskip 0.45cm}c@{\hskip 0.45cm}c} \hline
    \multirow{2}{4.2em}{Parameter} & \multirow{2}{5.3em}{Search~~range} & \multicolumn{3}{m{4.15cm}}{Intermediate resolution} & \multicolumn{3}{m{3.45cm}}{High resolution} \\
    & & Best fit & $1\sigma$ uncertainty & $3\sigma$ uncertainty & Best fit & $1\sigma$ uncertainty & $3\sigma$ uncertainty \\ \hline
    \multicolumn{8}{l}{\textbf{Mass model}} \\
    $\log\left(M_\mathrm{BH}/\mathrm{M}_\odot\right)$ & $8.70$ -- $9.95$ & \changes{$9.60$} & \changes{$\pm\,0.05$} & \changes{$-0.11$}, \changes{$+0.15$} & \changes{$9.55$} & $\pm\,0.02$ & \changes{$\pm\,0.07$} \\
    Inner stellar $M/L$ ($\mathrm{M}_\odot/\mathrm{L}_{\odot\mathrm{,F160W}}$) & $0.01$ -- $10$ & \changes{$3.04$} & \changes{$-0.28$, $+0.32$} & \changes{$-0.71$, $+1.1$} & \changes{$3.16$} & $\pm\,0.15$ & \changes{$-0.42$, $+0.41$}\\
    Outer stellar $M/L$ ($\mathrm{M}_\odot/\mathrm{L}_{\odot\mathrm{,F160W}}$) & $0.01$ -- $10$ & \changes{$2.41$} & \changes{$-0.17$, $+0.24$} & \changes{$-0.39$, $+0.60$} & \changes{$2.32$} & \changes{$-0.08$,$+0.10$} & \changes{$-0.25$, $+0.28$} \\ \hline
    \multicolumn{8}{l}{\textbf{Molecular gas disc}} \\
    Position angle (degree) & $112$ -- $172$ & $142.11$ & \changes{$-0.37$, $+0.34$} & \changes{$\pm\,1.1$} & $142.01$ & \changes{$-0.30$, $+0.29$} & \changes{$-0.90$, $+0.80$} \\
    Inclination (degree) & $26$ -- $89$ & \changes{$37.2$} & \changes{$-2.2$, $+2.0$} & \changes{$\pm\,5.3$} & $37.6$ & $\pm\,1.0$ & \changes{$-2.3$, $+2.8$} \\
    Velocity dispersion (\kms) & $0$ -- $25$ & \changes{$9.3$} & \changes{$\pm\,1.0$} & \changes{$-3.0$, $+3.4$} & \changes{$10.6$} & $\pm\,0.9$ & \changes{$-2.7$, $+2.8$} \\ \hline
    \multicolumn{8}{l}{\textbf{Nuisance parameters}} \\
    Integrated intensity (Jy~\kms) & $5$ -- $200$ & \changes{$74.8$} & \changes{$-4.8$, $+5.0$} & $-13$, $+15$ & \changes{$82.6$} & $\pm\,4.1$ & \changes{$-12$, $+11$} \\
    Kinematic centre $X$ offset (arcsec) & $-3.5$ -- $3.5$ & $-0.12$ & $\pm\,0.02$ & $\pm\,0.05$ & $-0.12$ & $\pm\,0.01$ & $\pm\,0.03$ \\
    Kinematic centre $Y$ offset (arcsec) & $-3.5$ -- $3.5$ & $0.00$ & $\pm\,0.02$ & \changes{$-0.05$, $+0.06$} & $0.00$ & $\pm\,0.01$ & $\pm\,0.03$ \\
    Systemic velocity (\kms) & $4890$ -- $5040$ & $4977.1$ & \changes{$-1.3$}, $+1.4$ & \changes{$-4.4$, $+4.1$} & $4971.9$ & $\pm\,1.0$ & \changes{$-2.8$, $+3.0$} \\ \hline
    \end{tabular}\\
    {\textsl{Notes.} The $X$ and $Y$ offsets are measured relative to the phase centre of the high-resolution data cube, $\mathrm{RA}=01^\mathrm{h}07^\mathrm{m}24\fs96$, $\mathrm{Dec.}=32\degree24\arcmin45\farcs21$ (J2000.0)}.
    \label{tab:results}
\end{table*}


\subsection{Results}
\label{subsec:results}

We fit the entire high-resolution data cube using a MCMC of $10^5$~steps, ensuring that the chain has converged and fully sampled the posterior distribution. The best-fitting parameters and their uncertainties are listed in the last three columns of Table~\ref{tab:results}. The one-dimensional marginalised posterior distribution of each non-nuisance parameter and the covariances between them are shown in Figure~\ref{fig:corner}. 

\begin{figure*}
    \centering
    \includegraphics[width=0.9\linewidth]{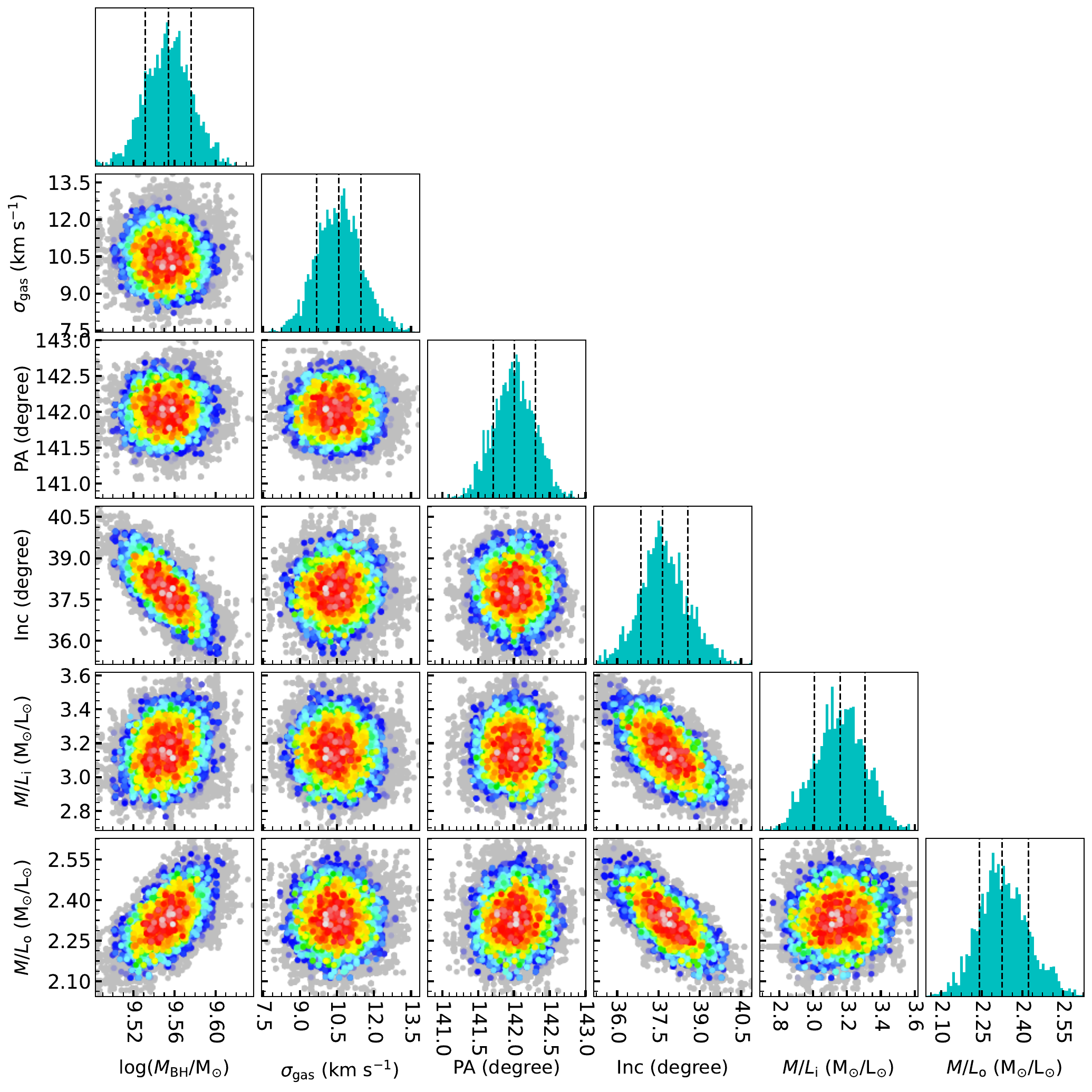}
    \caption{Corner plots showing the covariances between non-nuisance model parameters. The colours represent increasing confidence levels, from $68.3\%$ (red, $1\sigma$) to $99.7\%$ (grey, $3\sigma$). The histograms show the one-dimensional marginalised posterior distribution of each parameter; the dashed lines indicate the median and the $1\sigma$ confidence interval.
    }
    \label{fig:corner}
\end{figure*}

The best-fitting SMBH mass is \changes{$(3.58\pm0.19)\times10^9$~M$_\odot$}, in good agreement with that obtained from fitting the intermediate-resolution data cube. The SMBH mass precision has however improved from $\approx10\%$ to $\approx5\%$. The central panel of Figure~\ref{fig:pvd+model} shows the kinematic major-axis PVD of the best-fitting model overlaid on the observed PVD. The Keplerian rise (from the outside in) of the velocities up to $\approx635$~\kms\ due to the SMBH is well reproduced by our model. By contrast, a model with no SMBH (left panel of Figure~\ref{fig:pvd+model}) fails to account for the Keplerian rise. The best-fitting model of \citeauthor{North_2019} (\citeyear{North_2019}; right panel of Figure~\ref{fig:pvd+model}, with corrected kinematic centre and systemic velocity) reproduces the PVD nicely up to $\approx350$~\kms, the highest velocity detected by the intermediate-resolution observations, but it slightly overshoots the Keplerian rise beyond $\approx350$~\kms. Indeed, our best-fitting model has a reduced $\chi^2$ ($\chi_\mathrm{r}^2$) after rescaling of $1.59$ for the high-resolution data cube and $1.20$ for the intermediate-resolution data cube, better than the $\chi_\mathrm{r}^2$ of the best-fitting model of \citet{North_2019}, that are $1.97$ for the high-resolution data cube and $1.27$ for the intermediate-resolution data cube. This highlights the importance of high-resolution observations, that fully resolve the high-velocity material around the SMBH, to the precision and accuracy of SMBH mass measurements.

\begin{figure*}
    \captionsetup[subfigure]{labelformat=empty}
    \centering
    \subfloat[]{
    \includegraphics[width=0.332\linewidth]{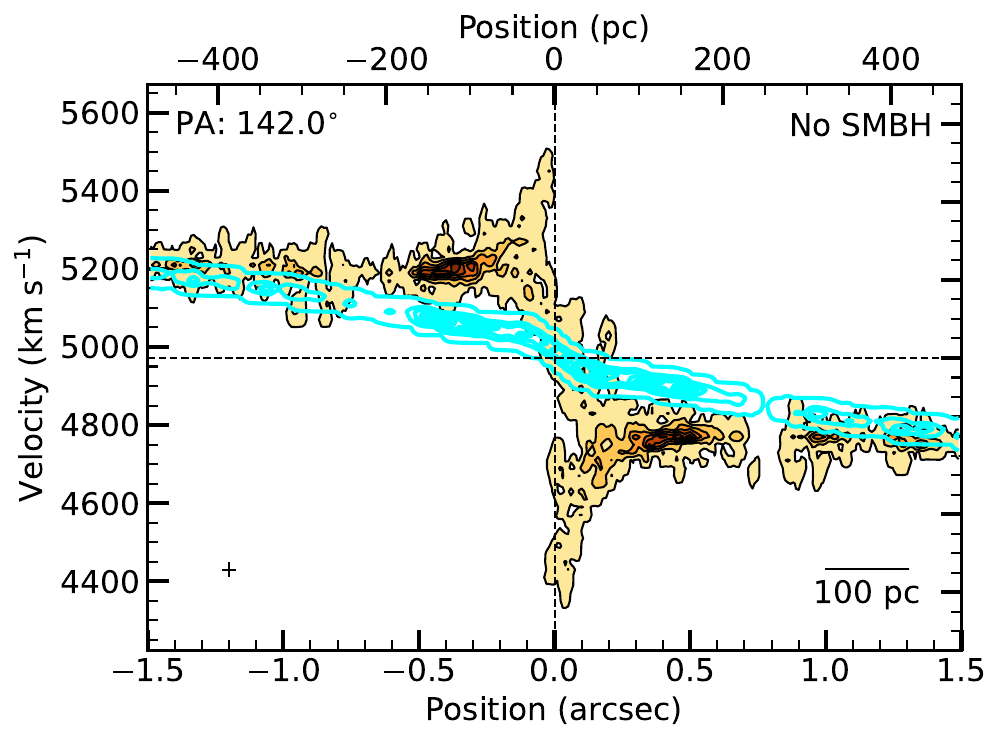}
    }
    \subfloat[]{
    \includegraphics[width=0.301\linewidth]{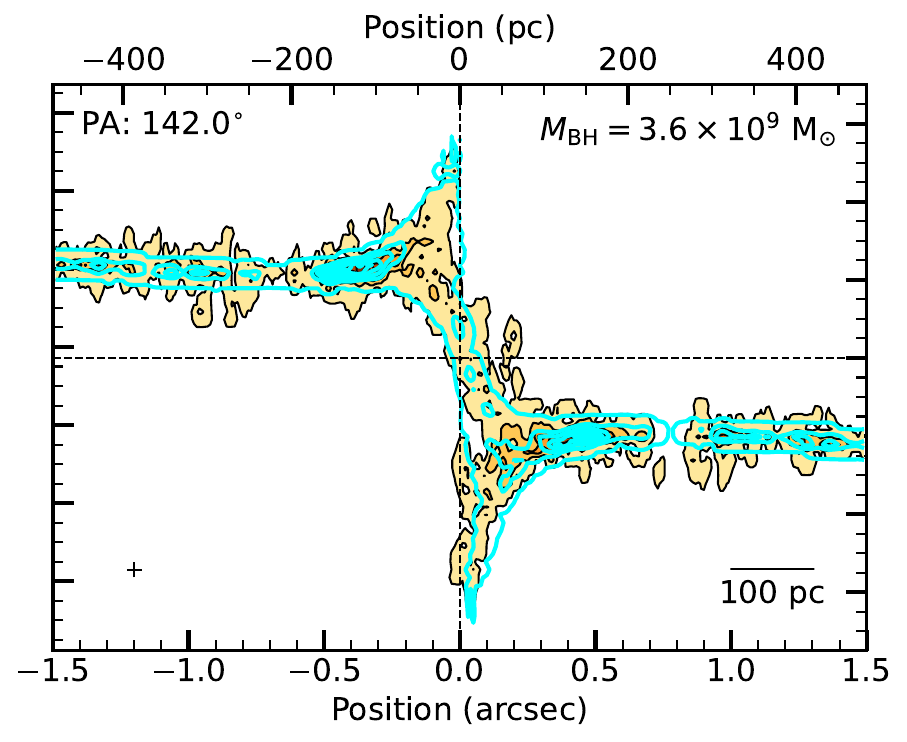}
    }
    \subfloat[]{
    \includegraphics[width=0.337\linewidth]{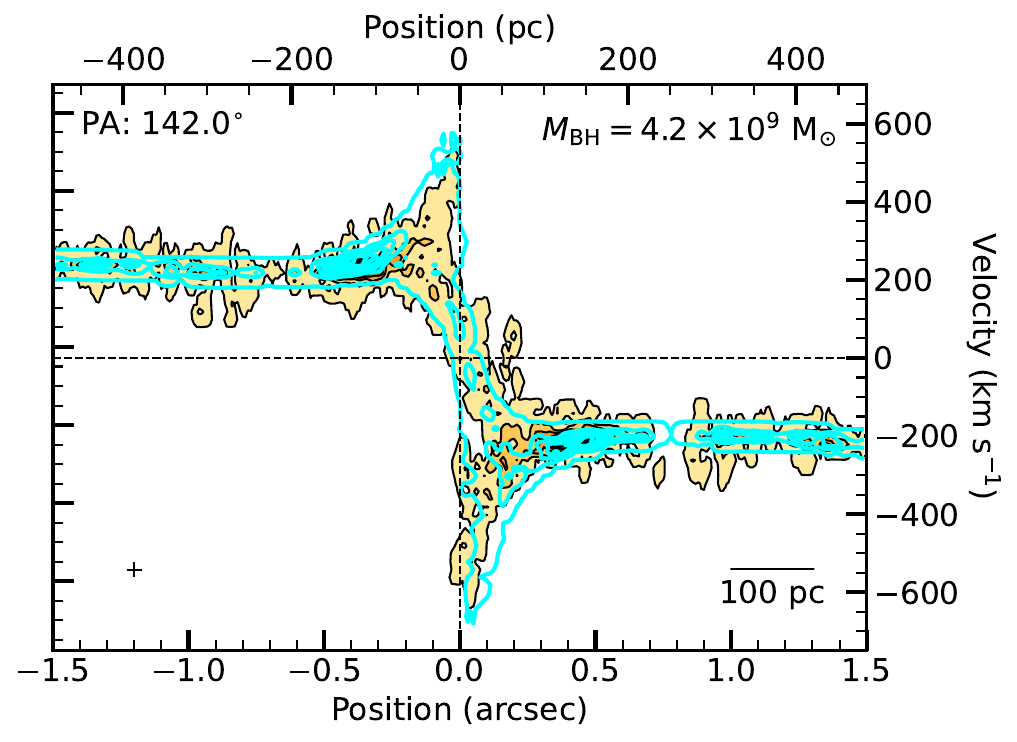}
    }
    \caption{Observed kinematic major-axis position-velocity diagram of NGC~383 (orange scale with black contours), overlaid with the PVDs of different models (cyan contours): no SMBH (left), best-fitting model from this work (centre) and best-fitting model from \citeauthor{North_2019} (\citeyear{North_2019}; right). Positions are measured relative to the best-fitting kinematic centre; velocities relative to the best-fitting systemic velocity along the right axis (see Section~\ref{subsec:results}). The cross in the bottom-left corner of each panel shows the synthesised beam FWHM along the kinematic major axis and the channel width. Our best-fitting model reproduces the material beyond $\approx350$~\kms\ better than the best-fitting model of \citet{North_2019}, demonstrating the importance of high-resolution observations to the precision and accuracy of SMBH mass measurements.}
    \label{fig:pvd+model}
\end{figure*}

We note that our best-fitting model does not reproduce the extended emission in the forbidden quadrants of the PVD, as it assumes an unwarped disc and purely circular orbits. We present models of this feature and show that it does not affect the best-fitting SMBH mass in Section~\ref{subsec:non-circ}. 

The stellar mass-to-light ratio of NGC~383 decreases from \changes{$3.16\pm0.15$}~$\mathrm{M}_\odot/\mathrm{L}_{\odot\mathrm{,F160W}}$ at the centre of the galaxy to \changes{$2.32\pm0.09$}~$\mathrm{M}_\odot/\mathrm{L}_{\odot\mathrm{,F160W}}$ at the outer edge of the molecular gas disc ($R=3\farcs5$). Both the inner and the outer $M/L$ agree with those of the intermediate-resolution fit. 
However, the systemic velocity is lower by $\approx2.9\sigma$, potentially because of the larger channel width ($20$~\kms) of the high-resolution data cube.


\section{Discussion} \label{sec:discussion}

\subsection{Warps and/or non-circular motions}
\label{subsec:non-circ}

A potential source of systematic error in our model is the simplification that the molecular gas moves in circular orbits in an unwarped disc. Contrary to this assumption, the high-resolution velocity map (right panel of Figure \ref{fig:moments}) reveals a twist of the isovelocity contours within the central $\approx0\farcs3$ in radius, suggesting a PA warp and/or non-circular motions. A distortion of the velocity field is also suggested by the forbidden quadrant emission in the kinematic major-axis PVD (Figure \ref{fig:pvd}). We thus also created a kinematic minor-axis PVD of the data, shown in Figure~\ref{fig:natural_minor_pvd}. For a rotating disc with no warp nor non-circular motion, the minor-axis PVD traces the zero-velocity contour as well as some emission at non-zero velocities due to the velocity dispersion, beam smearing and the finite width of the pseudo slit, that should be perfectly symmetric in all four quadrants. By contrast, the minor-axis PVD of NGC~383 appears asymmetric, with more positive-velocity emission near $-0\farcs2$ and more negative-velocity emissions near $+0\farcs2$, consistent with the shape of the isovelocity contour twist.

\begin{figure*}
    \captionsetup[subfigure]{labelformat=empty}
    \centering
    \subfloat[]{
    \includegraphics[width=0.332\linewidth]{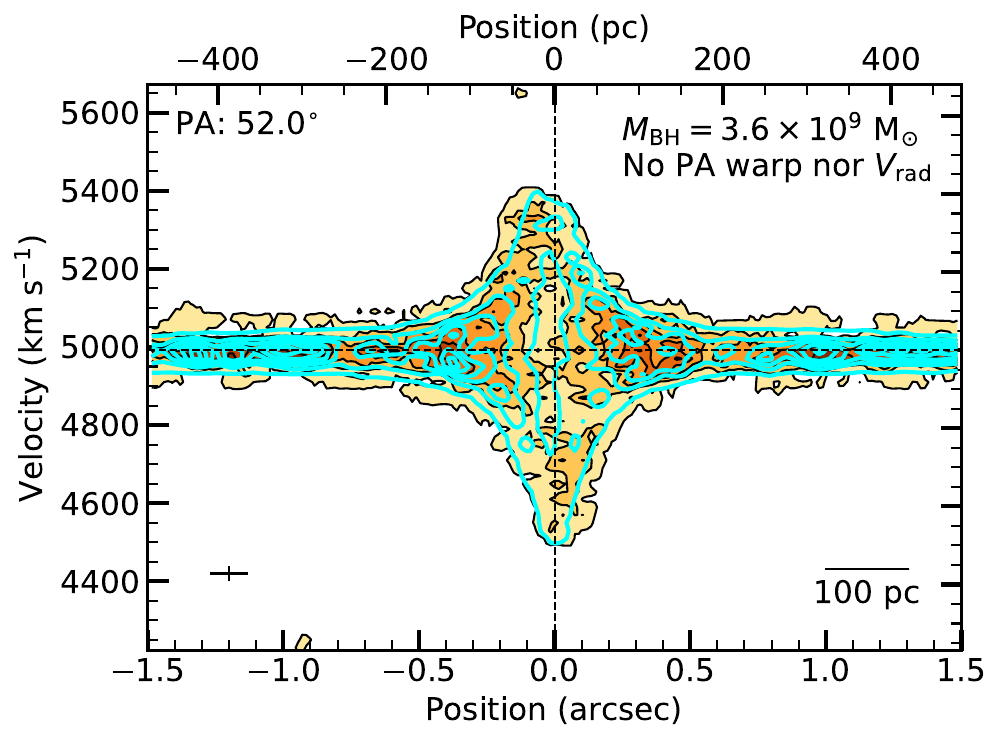}
    }
    \subfloat[]{
    \includegraphics[width=0.301\linewidth]{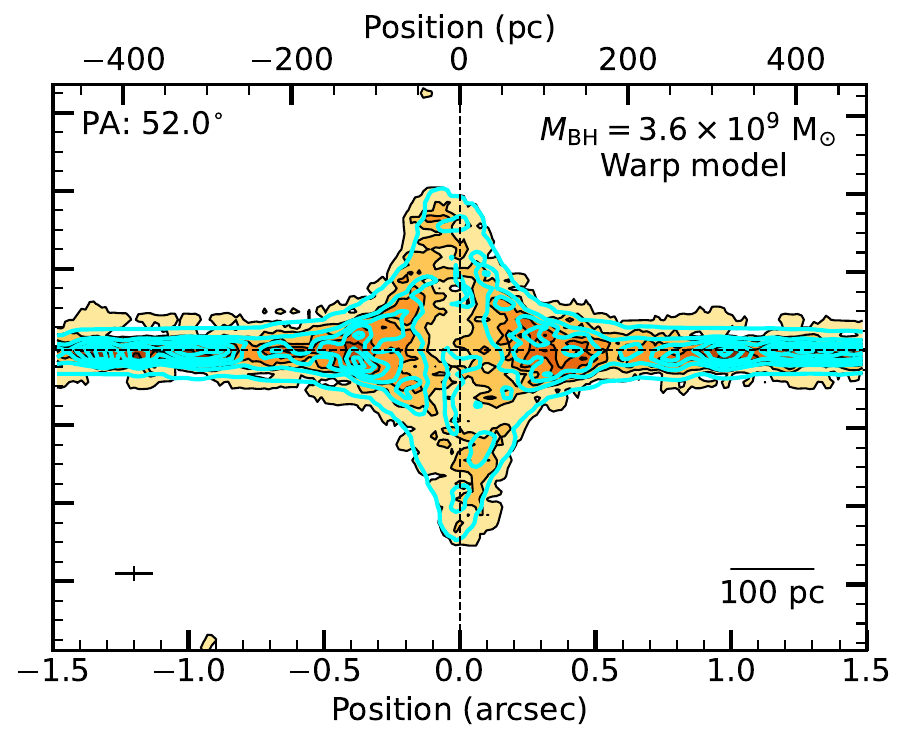}
    }
    \subfloat[]{
    \includegraphics[width=0.337\linewidth]{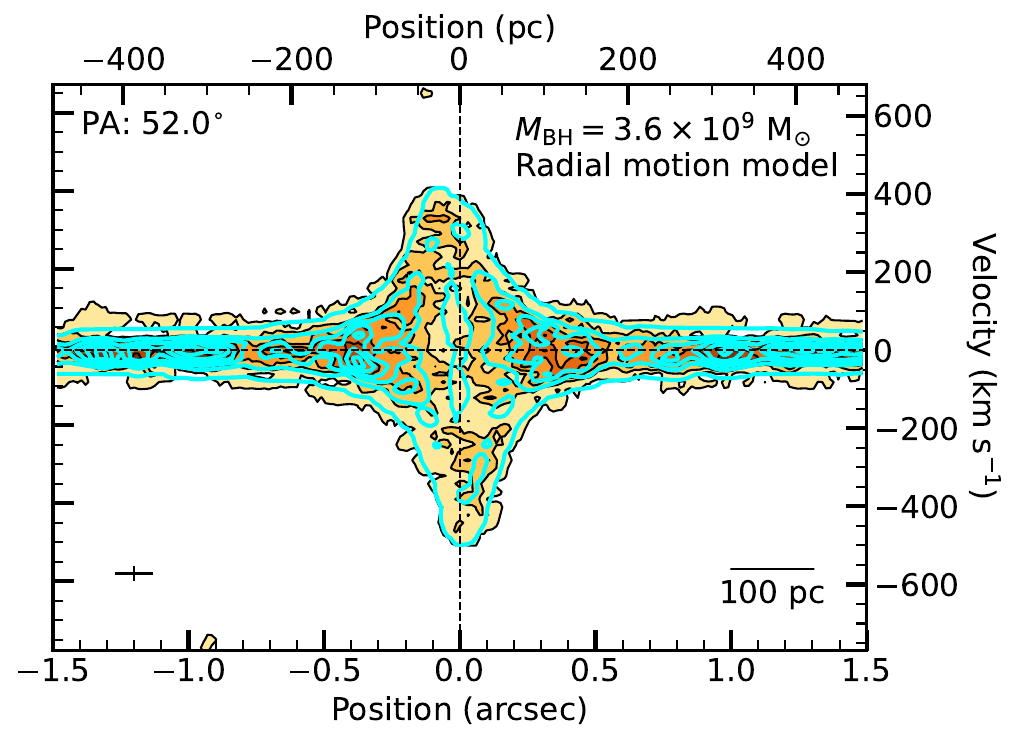}
    }
    \caption{Same as Figure \ref{fig:pvd+model}, but showing the kinematic minor-axis PVD of NGC~383 imaged with natural weighting, overlaid with the PVDs of the best-fitting model with no warp nor radial motion (left), a warp (centre) and radial motions (right). Although the warp model suggests a different minor-axis PA in the innermost region, we adopt the same large-scale PA as other models to compare them. The asymmetry of the observed minor-axis PVD within the central $\approx0\farcs2$ in radius is consistent with the observed twist of the zero-velocity contour. Both the warp and the radial motion model partly reproduce the asymmetry, while the model with no warp nor radial motion has a symmetric minor-axis PVD by construction.
    }
    \label{fig:natural_minor_pvd}
\end{figure*}

Despite the many features that suggest a PA warp and/or non-circular motions, the low $S/N$ and the relatively small number of pixels in the region of the velocity twist make it challenging to model those features using the current data cube. Instead, we re-image the data with Briggs weighting and a robust parameter of $2.0$ (equivalent to natural weighting), to maximize the $S/N$. The resulting data cube has a synthesised beam FWHM of $0\farcs093\times0\farcs062$ ($\approx30\times20$~pc$^2$) with a PA of $12\fdg9$, sufficient to resolve the velocity twist. The twist is more prominent in the velocity map of the new data cube (see the left panels of Figure~\ref{fig:velo_resi}), thanks to the improved $S/N$ (new cube dynamic range, i.e.\ peak $S/N$,  of $\approx22$). We thus only use this naturally-weighted data cube for the investigations in this sub-section.

\begin{figure*}
\captionsetup[subfigure]{labelformat=empty}
    \centering
    \subfloat[]{
    \includegraphics[width=0.32\linewidth]{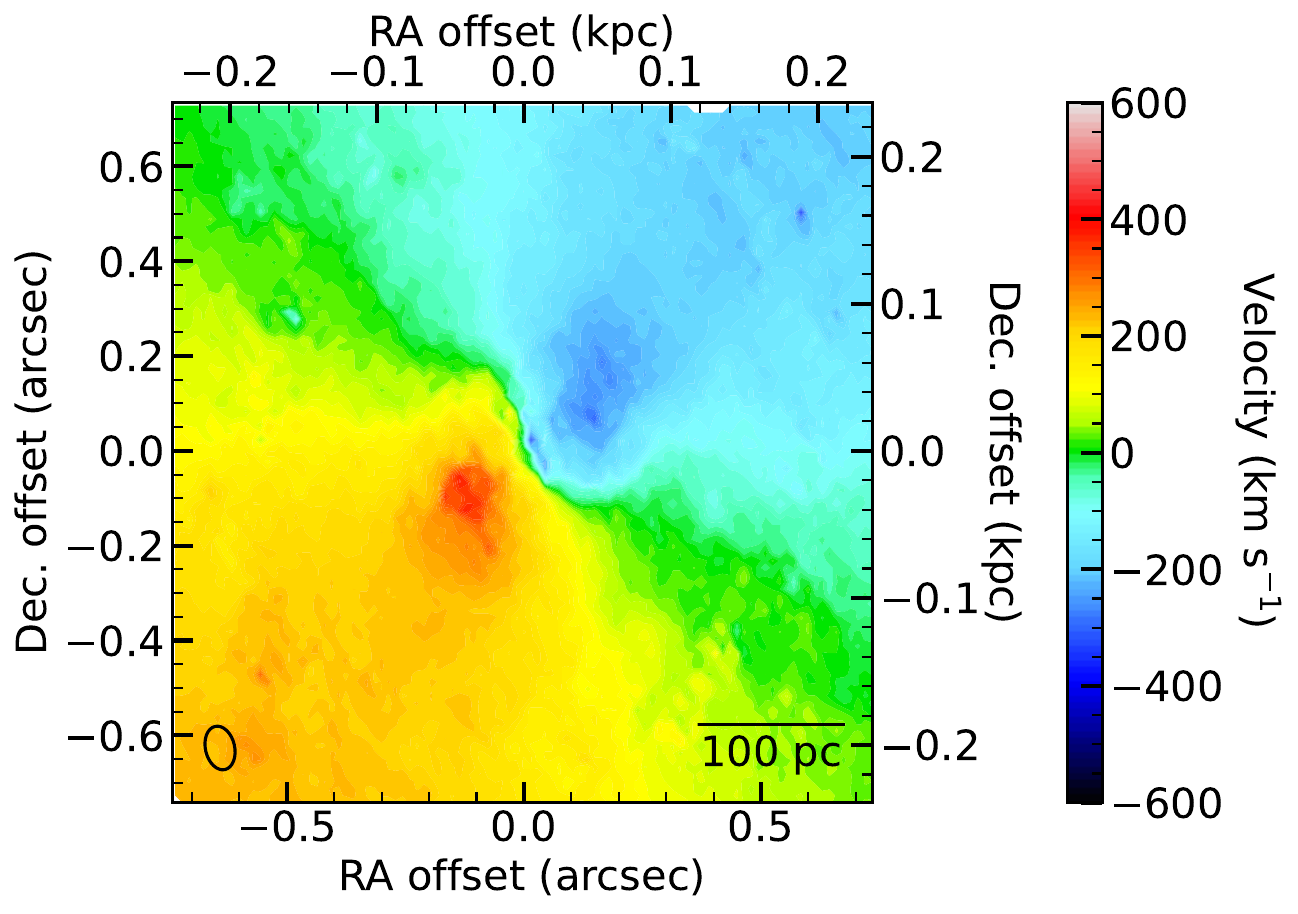}
    }
    \subfloat[]{
    \includegraphics[width=0.32\linewidth]{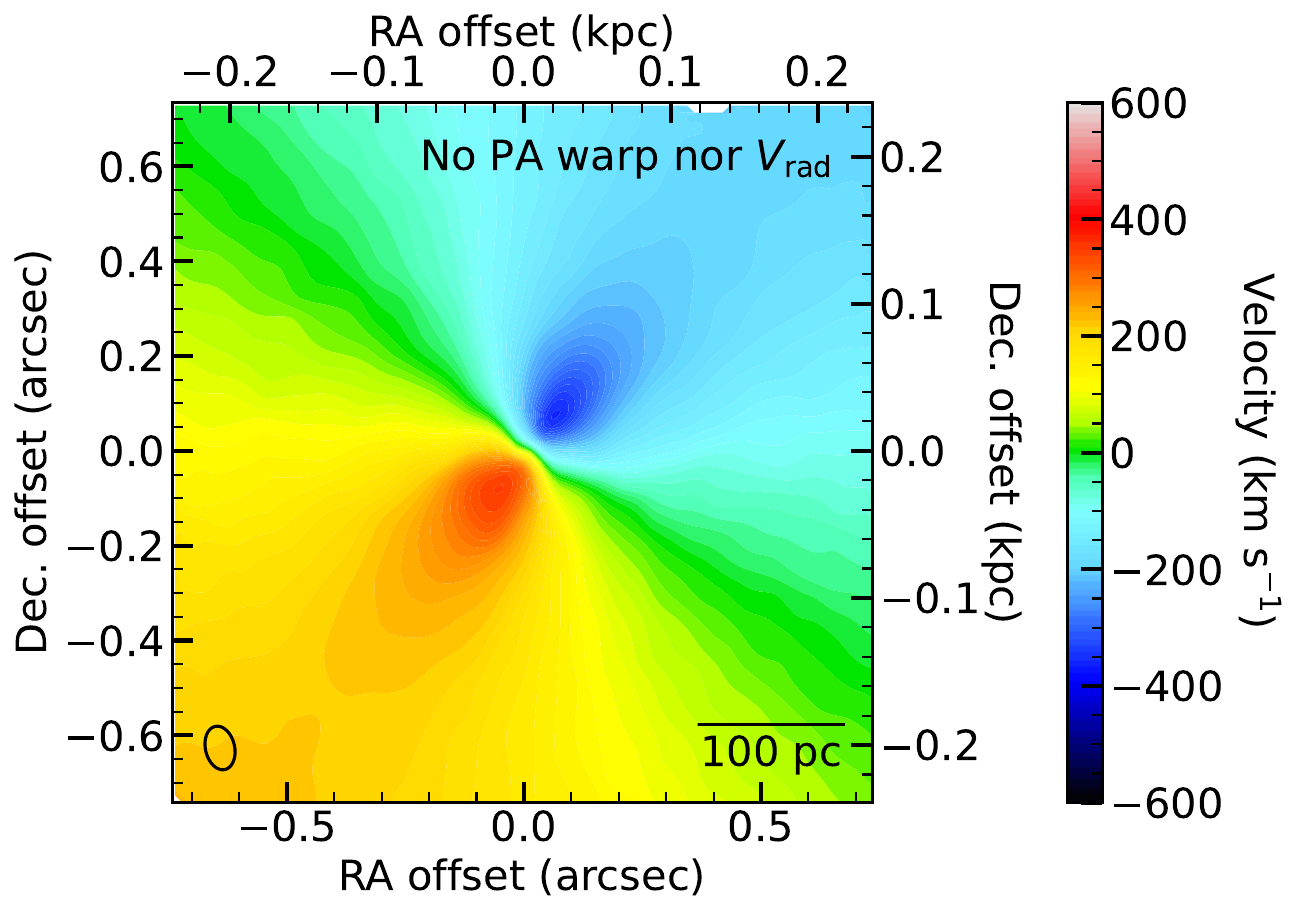}
    }
    \subfloat[]{
    \includegraphics[width=0.32\linewidth]{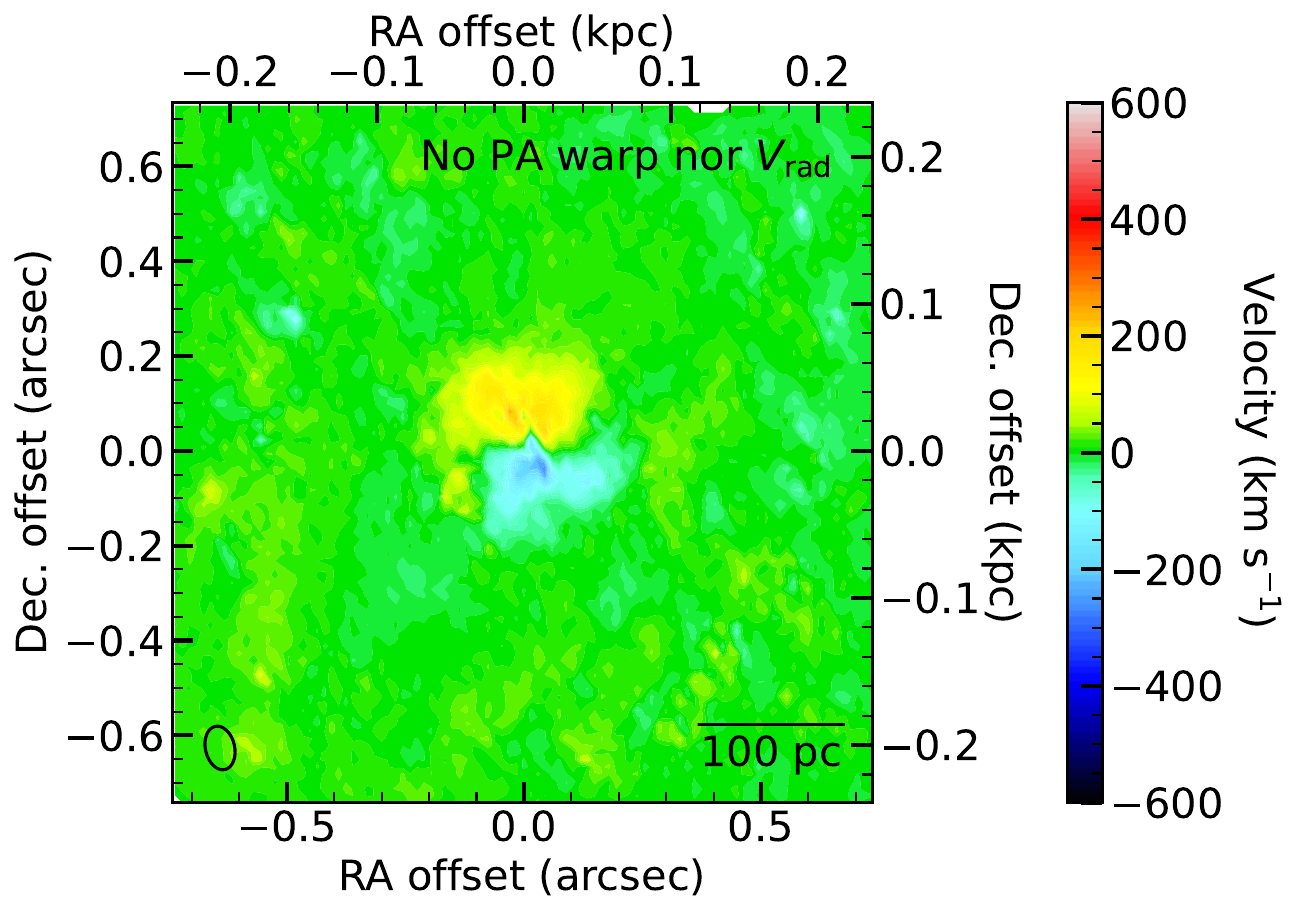}
    } \\
    \subfloat[]{
    \includegraphics[width=0.32\linewidth]{Figures/NGC0383_natural_velo_data.pdf}
    }
    \subfloat[]{
    \includegraphics[width=0.32\linewidth]{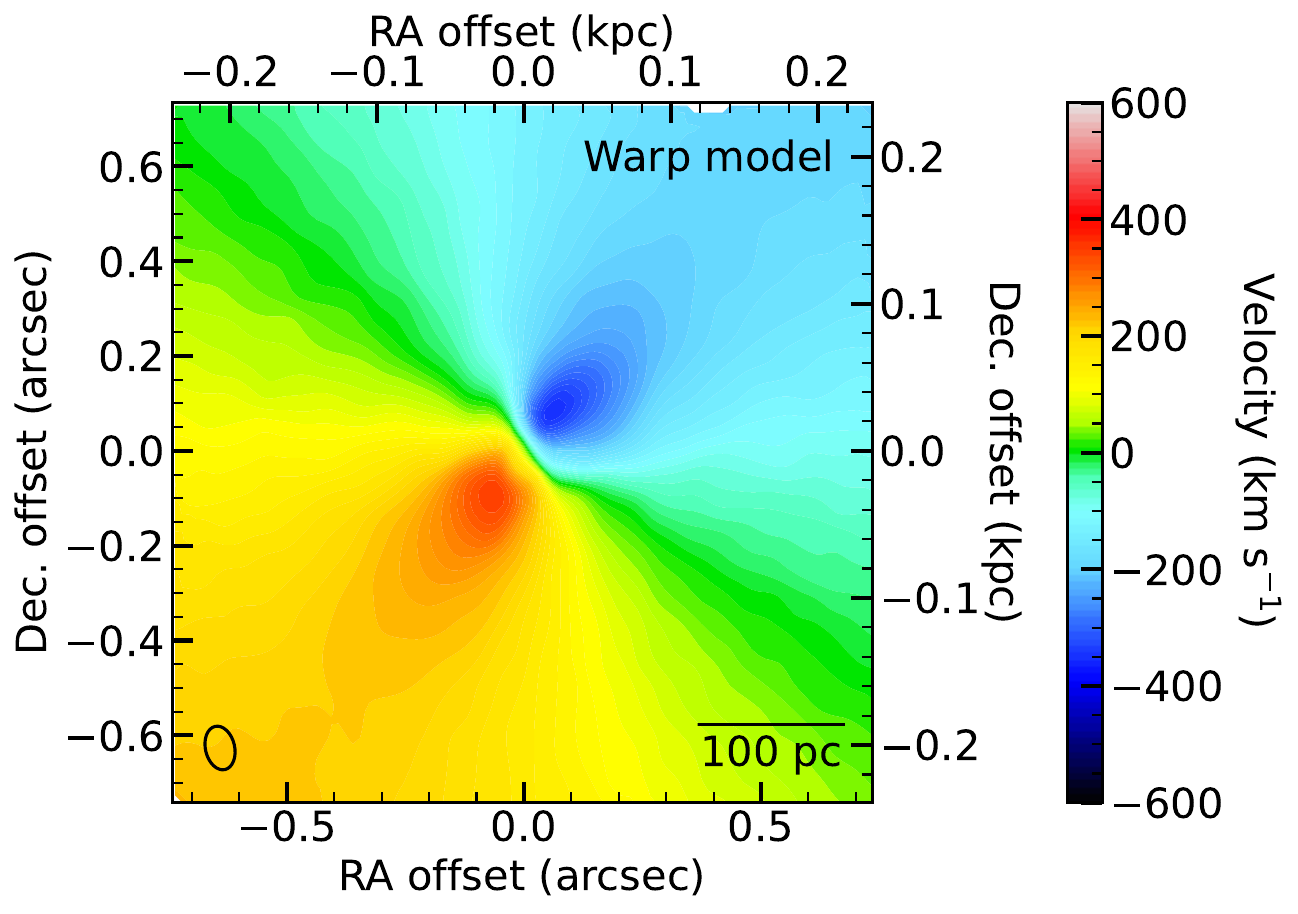}
    }
    \subfloat[]{
    \includegraphics[width=0.32\linewidth]{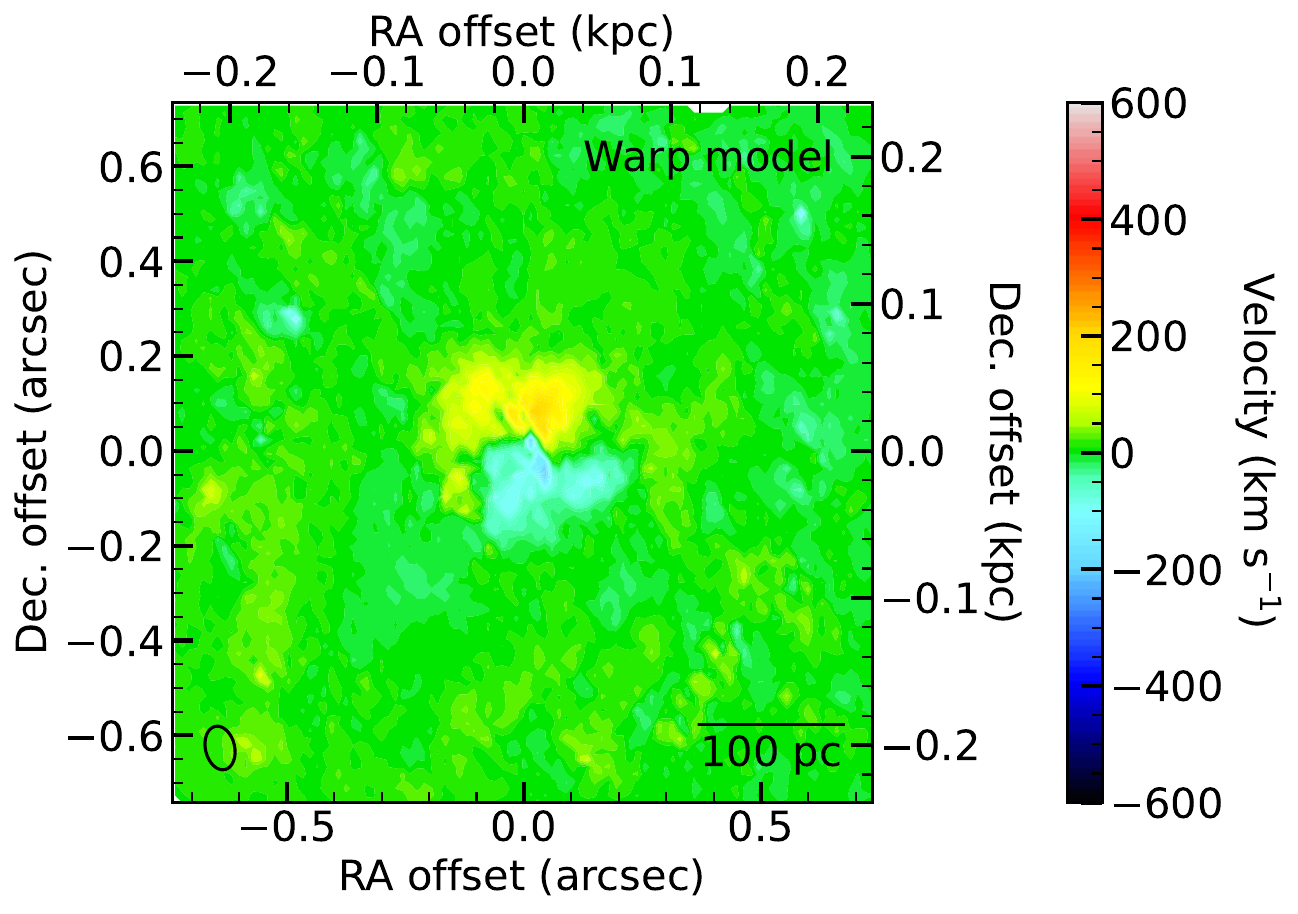}
    } \\
    \subfloat[]{
    \includegraphics[width=0.32\linewidth]{Figures/NGC0383_natural_velo_data.pdf}
    }
    \subfloat[]{
    \includegraphics[width=0.32\linewidth]{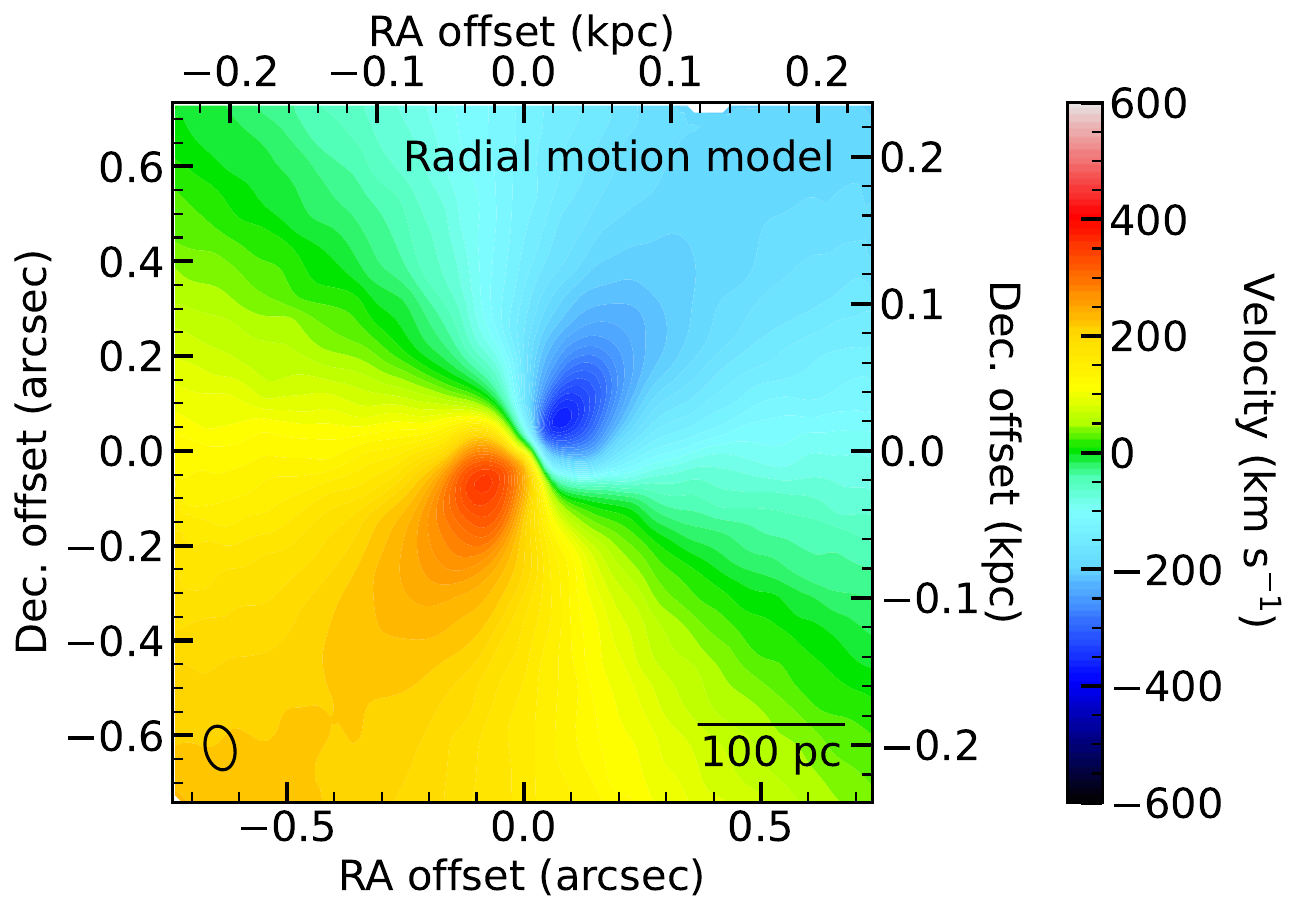}
    }
    \subfloat[]{
    \includegraphics[width=0.32\linewidth]{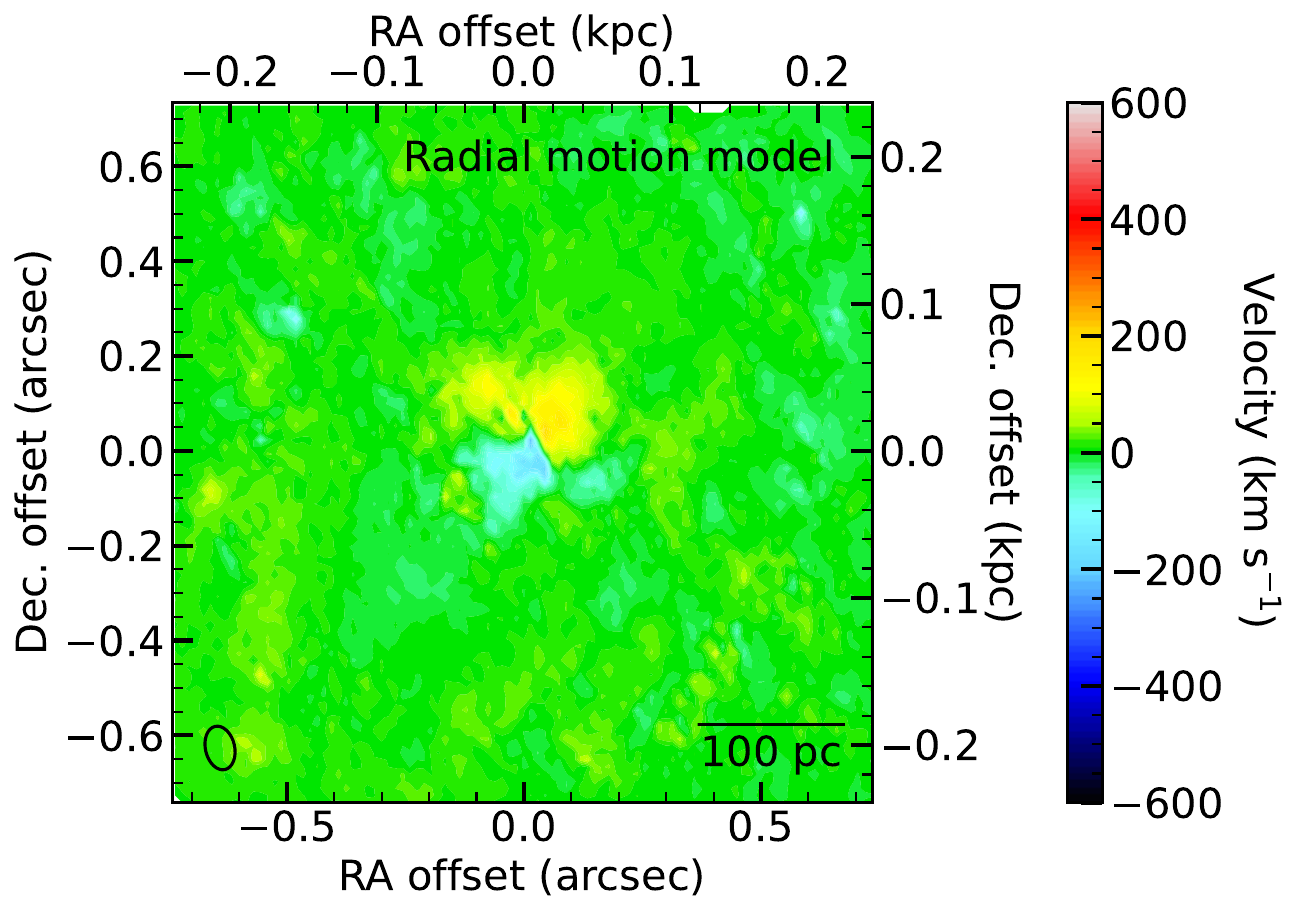}
    } \\
    
    \caption{
    Left panels: first-moment (intensity-weighted mean line-of-sight velocity) map of the NGC~383 data cube imaged with natural weighting, showing only the central $1\farcs5\times1\farcs5$. The central velocity twist is more prominent as natural weighting maximises the $S/N$. Middle panels: first-moment maps of the best-fitting dynamical model with no PA warp nor non-circular motion (top), the best-fitting warp model (middle) and the best-fitting radial motion model (bottom). Right panels: corresponding first-moment residual maps ($\mathrm{data}-\mathrm{model}$). The synthesised beam ($0\farcs093\times0\farcs062$) is shown as a black open ellipse in the bottom-left corner of each panel, while a $100$~pc scale bar is shown in the bottom-right corner of each panel. Positions are measured relative to the best-fitting kinematic centre; velocities relative to the best-fitting systemic velocity (see Section~\ref{subsec:results}). The model with no warp nor non-circular motion produces no velocity twists, leaving residuals up to $\pm280$~\kms\ in the inner parts. Both the warp and the radial motion models partially reproduce the central velocity twist, but large velocity residuals ($\approx210$ and $\approx220$~\kms, respectively) remain at the centre.
    }
    \label{fig:velo_resi}
\end{figure*}

\subsubsection{Tilted-ring models}

To create a non-parametric model of a PA warp and/or non-circular motions, we first use the \texttt{3DFIT} task of the $^{\textsc{3D}}\textsc{Barolo}$ package\footnote{Version 1.7 from \url{https://bbarolo.readthedocs.io/en/latest/}} \citep{DiTeodroro_2015} to fit a tilted-ring model to the data cube. We divide the central $1\farcs2$ in radius of the cube into $20$ rings, each with a width of $0\farcs06$ (approximately the FWHM of the synthesised beam's minor axis), fully covering and sufficiently sampling the region of the velocity twist. By default, each ring is characterised by ten parameters: $x$ and $y$ coordinates of the ring's centre $(x_0,y_0)$, systemic velocity ($V_\mathrm{sys}$), position angle (PA), inclination (Inc), gas surface density ($\Sigma_\mathrm{gas}$), vertical thickness ($z_0$), rotation velocity ($V_\mathrm{rot}$), radial velocity ($V_\mathrm{rad}$) and velocity dispersion ($\sigma_\mathrm{gas}$). We normalise the gas surface density using the azimuthal average of the observed $^{12}$CO(2-1) intensity in each ring and fix $z_0$ to be zero (infinitely thin disc approximation), leaving only eight free parameters. We set "TWOSTAGE=true" so that all parameters vary freely in the first fitting stage, but the geometric parameters $x_0$, $y_0$ and $V_\mathrm{sys}$ of each ring are regularised and kept fixed (by default to the median of the $20$ rings) in the second fitting stage.

As disc warps and non-circular motions are often degenerate, causing similar distortions of the isovelocity contours, we perform two separate fits. In the first fit (warp model), the PA and Inc of each ring are regularised by Bezier interpolation in the second fitting stage, allowing them to vary across rings (i.e.\ create a warp), but the radial velocity is fixed to $V_\mathrm{rad}=0$. In the second fit (radial motion model), the PA and Inc are fixed to the median across the rings in the second fitting stage, but the radial velocity $V_\mathrm{rad}$ is left free, providing a pure radial motion model. In both fits, we only consider pixels within a boolean mask, constructed similarly to the mask used to create the moment maps in Section~\ref{subsec:line_em}, except that we clip at $1$~$\sigma_\mathrm{RMS}$ of the naturally-weighted data cube. All other parameters of the \texttt{3DFIT} task are set to their defaults.

For the warp model, the best-fitting ring parameters yielded by the second fitting stage suggest that the velocity field at the centre of NGC~383 is consistent with an abrupt change of PA from $142\degree$ in the outer region to $\approx110\degree$ in the inner region near $R=0\farcs2$. However, the PA of the innermost region is highly uncertain and is sensitive to the initial guess. Any PA between $\approx100\degree$ and $\approx130\degree$ yields a similar RMS of $\approx30$~\kms\ in the first-moment (i.e.\ intensity-weighted mean line-of-sight velocity) residuals within the central $1\farcs2$ in radius. For reference, a model with no PA warp nor radial motion has a RMS residual velocity of $\approx50$~\kms\ in the same region. 
For the radial motion model, Figure~\ref{fig:BBarolo_RV} shows the radial profile of the radial velocities yielded by the second fitting stage. $V_\mathrm{rad}$ is inconsistent with zero (to $1\sigma$) at $R\leq0\farcs21$ only, revealing a roughly linear trend of increasing radial velocity towards the centre. The RMS of the residual velocities is $\approx20$~\kms. The $^{\textsc{3D}}\textsc{Barolo}$ tilted-ring fits therefore provide mildly more compelling evidence for radial motions than for a PA warp. To determine whether the radial motions represent an inflow or an outflow, one has to know which side of the gas disc is closer to the observer. As the dust lanes in the galaxy's \textsl{HST} image appear only on the southwestern side of the galaxy \citep{North_2019}, the southwestern side is likely the near side, and the gas disc is likely rotating in a clockwise direction in the plane of the sky. The radial motions (if confirmed) therefore likely represent an outflow.

\begin{figure}
    \centering
    \includegraphics[width=\linewidth]{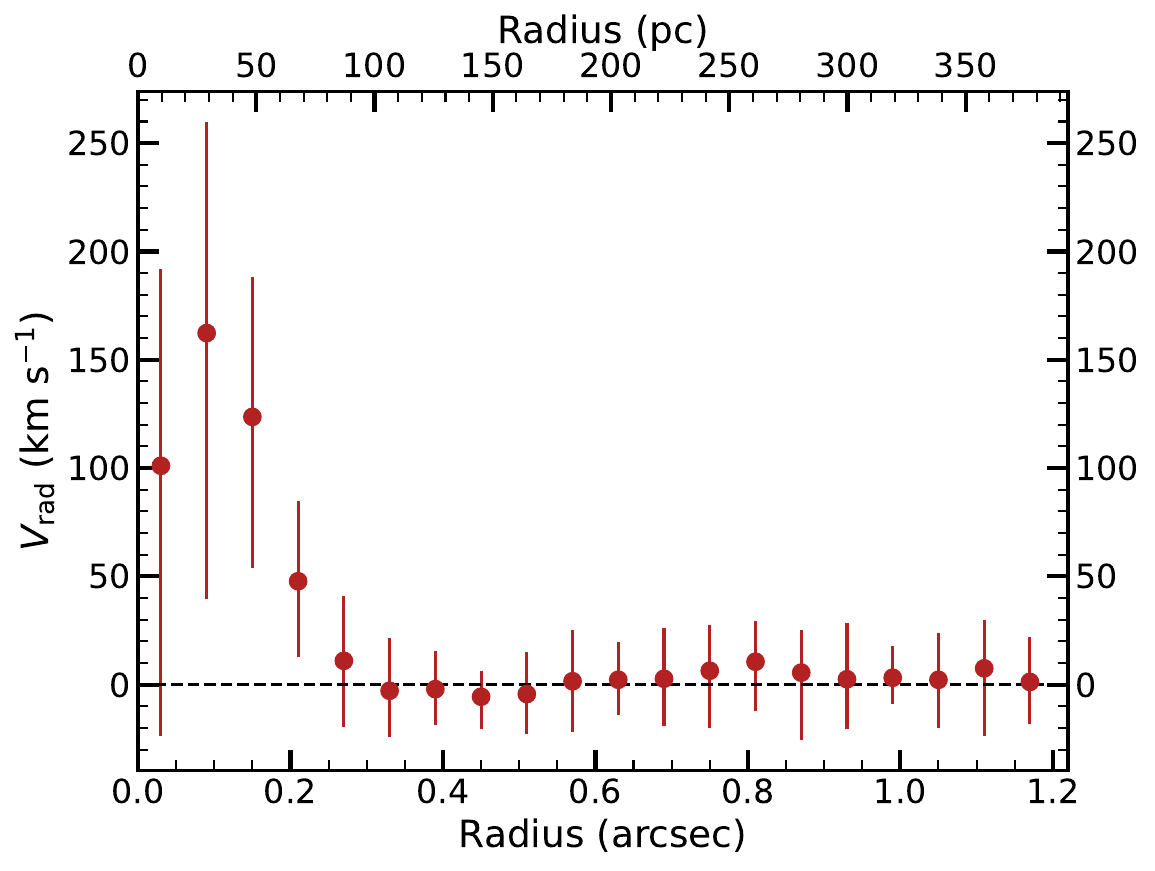}
    \caption{Radial profile of the best-fitting radial velocities (with $1\sigma$ error bars), from our $^{\textsc{3D}}\textsc{Barolo}$ radial motion model of NGC~383. The radial velocity is inconsistent with zero (to $1\sigma$) at $R\leq0\farcs21$ only, revealing evidence of radial motions in the innermost region of NGC~383.
    }
    \label{fig:BBarolo_RV}
\end{figure}

\subsubsection{Parametric models}

To obtain parametric models of the PA warp and radial motions and fully sample the posterior distributions of the parameters, we repeat the above analysis with \textsc{KinMS}. While $^{\textsc{3D}}\textsc{Barolo}$ allows the kinematics of each ring to vary freely without considering the underlying mass distribution, the circular velocity curve of \textsc{KinMS} is determined by the mass profile. \textsc{KinMS} thus allows us to check whether the best-fitting SMBH mass changes after incorporating a PA warp or non-circular motions in the kinematic model. Motivated by the $^{\textsc{3D}}\textsc{Barolo}$ results, we thus use \textsc{KinMS} to fit a warp involving an abrupt PA change at a particular radius but no radial motion (warp model) and to fit radial motions with a linear variation of the radial velocities with radius but no warp (radial motion model). The warp model replaces the PA in the model of Section~\ref{sec:modelling} with three new free parameters: the PA in the inner region ($\mathrm{PA}_\mathrm{inner}$), the PA in the outer region ($\mathrm{PA}_\mathrm{outer}$) and the radius at which the PA changes abruptly ($R_\mathrm{change}$). The radial motion model has two additional parameters: the radial velocity at $R=0$ ($V_\mathrm{rad,0}$) and the radius beyond which the radial velocity is $0$ ($R_\mathrm{cutoff}$). Other parameters are identical to the ones listed in Section~\ref{sec:modelling} for the model with no warp nor radial motion (hereafter "regular disc model").

The best-fitting warp model yields an inner PA \changes{$\mathrm{PA}_\mathrm{inner}=117\degree_{-31\degree}^{+25\degree}$} and an outer PA \changes{$\mathrm{PA}_\mathrm{outer}=142\fdg08\pm0\fdg13$}. The PA changes abruptly at $R_\mathrm{change}=0\farcs18\pm0\farcs05$. The remaining best-fitting parameters, including the SMBH mass, are almost identical to those of the regular disc model. Because the inner PA is consistent with the outer PA within $1\sigma$, the warp model is not strongly favoured over models with no PA warp. Table~\ref{tab:model_stats} compares the statistics of the best-fitting warp model to those of the best-fitting regular disc model. Although the warp model has a slightly lower $\chi^2_\mathrm{r}$, it has two more free parameters. Considering instead the Bayesian information criterion (BIC), defined as $\mathrm{BIC}\equiv k\ln N-2\ln P$, where $k$ is the number of free parameters, $N$ the number of constraints, and $-2\ln P$ is the $\chi^2$ defined in Section~\ref{sec:modelling}, the warp model is actually less preferred than the regular disc model. The top panels and the middle panels of Figure~\ref{fig:velo_resi} show the central $1\farcs5\times1\farcs5$ of first-moment map of the regular disc model and of the warp model, respectively, the data and the residuals ($\mathrm{data}-\mathrm{model}$).  The mask used to create the first-moment maps here is again similar to that used in Section~\ref{subsec:line_em}, but we clip at $0.65$~$\sigma_\mathrm{RMS}$ of the naturally-weighted data cube. As the regular disc model does not produce any velocity twist, velocity residuals up to $\pm280$~\kms\ remain in the central region. By contrast, the warp model produces a mild velocity twist and reduces the velocity residuals. However, it does not fully reproduce the observed twist and leaves residuals up to $\approx 210$~\kms\ at the centre. We have also tested other parametric models of a PA warp (e.g.\ linear warp with radius), but none yields a better BIC than the current model. This strongly suggests that a PA warp does not satisfactorily explain the twist observed in the velocity field.

\begin{table}
    \centering
    \caption{Statistics of the models discussed.}
    \begin{tabular}{lccccc} \hline
    Model & Max $\abs{V_\mathrm{res}}$ & RMS $V_\mathrm{res}$ & $\chi^2_\mathrm{r}$ & $k$ & BIC \\
    & (\kms) & (\kms) & & & \\
    (1) & (2) & (3) & (4) & (5) & (6) \\ \hline
    Regular disc & $281$ & $26$ & $2.080$ & $10$ & $12043$ \\
    Warp & $207$ & $24$ & $2.075$ & $12$ & $12063$ \\
    Radial motion & $224$ & $23$ & $2.072$ & $12$ & $12060$ \\ \hline
    \end{tabular}
    \label{tab:model_stats}\\
    {{\sl Notes.} (1) Model. (2) Maximum velocity residual (central $1\farcs5\times1\farcs5$). (3) RMS velocity residual (central $1\farcs5\times1\farcs5$). (4) Reduced $\chi^2$ statistic (entire cube). (5) Number of free parameters. (6) Bayesian information criterion.}
\end{table}

It is also worth noting that if a PA warp from $142\degree$ in the outer CO disc to $\approx110\degree$ in the inner CO disc is present in NGC~383, then the rotation axis of the inner CO disc would align better with the radio jet (with a position angle $\mathrm{PA}_\mathrm{jet}=-19.7\degree\pm1.0\degree$; \citealt{Laing_2014}) than that of the outer CO disc. Although recent studies \citep[e.g.][]{Ruffa_2020} suggest that jets and $100$-pc-scale CO discs are not preferentially aligned, the relative orientations of jets and the rotation axes of sub-pc maser discs are usually within $\approx30\degree$ \citep{Kamali_2019}. As our high-resolution observations of NGC~383 probe the same physical scale (in the unit of the Schwarzschild radius) as maser observations \citep{Zhang_2024}, the PA warp could be an indication that our observations start to reveal CO gas more closely related to the central accretion disc that powers the AGN jet than earlier CO observations.

The best-fitting radial motion model suggests a radial velocity linearly decreasing from \changes{$V_\mathrm{rad}=294_{-181}^{+136}$}~\kms\ at $R=0$ to $V_\mathrm{rad}=0$ at \changes{$R_\mathrm{cutoff}=0\farcs28_{-0.07}^{+0.05}$}. These parameters are consistent with the $^{\textsc{3D}}\textsc{Barolo}$ tilted-ring fits (Figure~\ref{fig:BBarolo_RV}) but have large uncertainties. We thus only have tentative evidence of radial motions, with $V_\mathrm{rad,0}$ only \changes{$\approx1.6\sigma$} different from zero. The remaining best-fitting parameters, including the SMBH mass, are all in agreement with those of the regular disc model. Table~\ref{tab:model_stats} lists the statistics of the radial motion model, while the bottom panels of Figure~\ref{fig:velo_resi} shows the first-moment maps of the model and data, and the associated velocity residual map. The radial motion model has the smallest $\chi^2_\mathrm{r}$ of the three models, but it is still not preferred over the regular disc model when considering the BIC. It yields the smallest RMS velocity residual in the central region, but large residuals up to $\approx220$~\kms\ remain, suggesting non-circular motions more complicated than pure radial motions.

Finally, we compare the kinematic major-axis and minor-axis PVDs of the three models to those of the data in Figures~\ref{fig:natural_major_pvd} and \ref{fig:natural_minor_pvd}, respectively. Although the warp model suggests a different major-axis PA in the innermost region, we adopt the same large-scale PA as other models to compare them. Both the warp and the radial motion models partly but not fully reproduce the forbidden quadrant emission of the major-axis PVD and the asymmetry of the minor-axis PVD, suggesting again that neither model fully explains the observed distortions of the velocity field.

\changes{We note that we have also tested parametric models combining a PA warp and radial motions. However, the models' PA warp and radial motion parameters could not be simultaneously constrained, leading to unconverged MCMC chains, nor could the combined models significantly reduce the velocity residuals, again because the data do not have sufficient $S/N$ to break the degeneracy between a PA warp and a non-zero $V_\mathrm{rad}$.}

\begin{figure*}
    \captionsetup[subfigure]{labelformat=empty}
    \centering
    \subfloat[]{
    \includegraphics[width=0.332\linewidth]{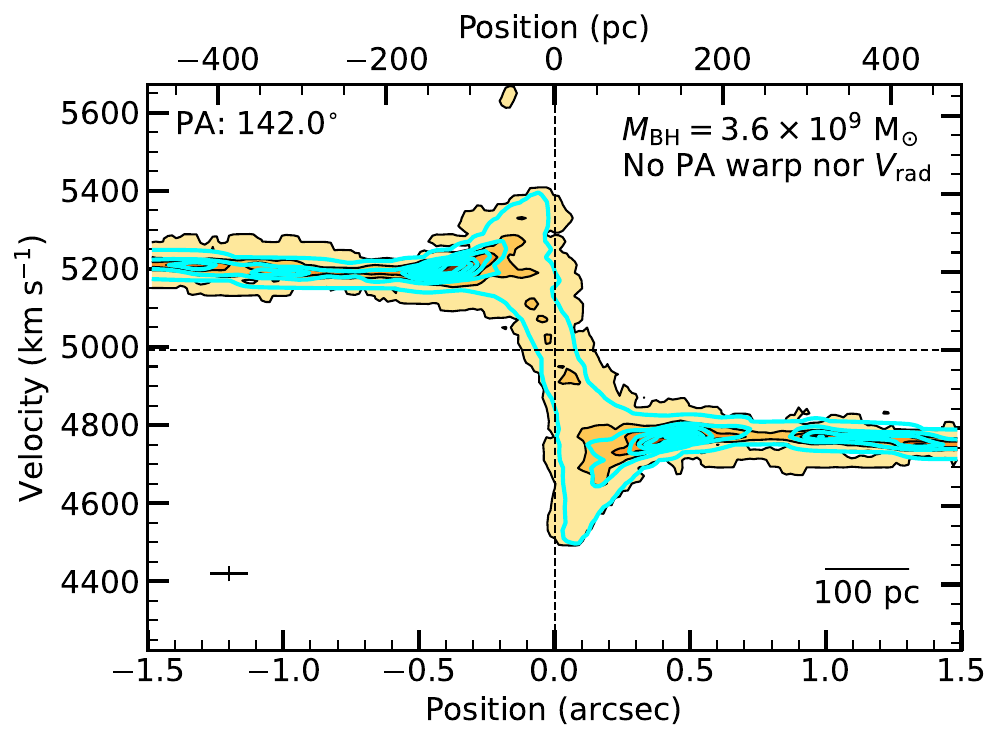}
    }
    \subfloat[]{
    \includegraphics[width=0.301\linewidth]{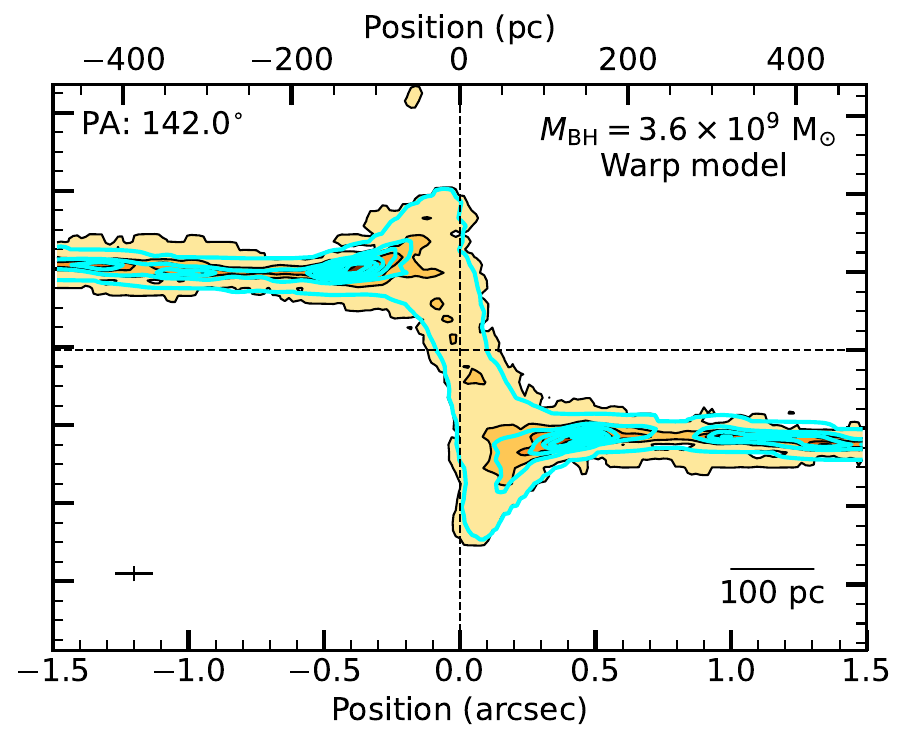}
    }
    \subfloat[]{
    \includegraphics[width=0.337\linewidth]{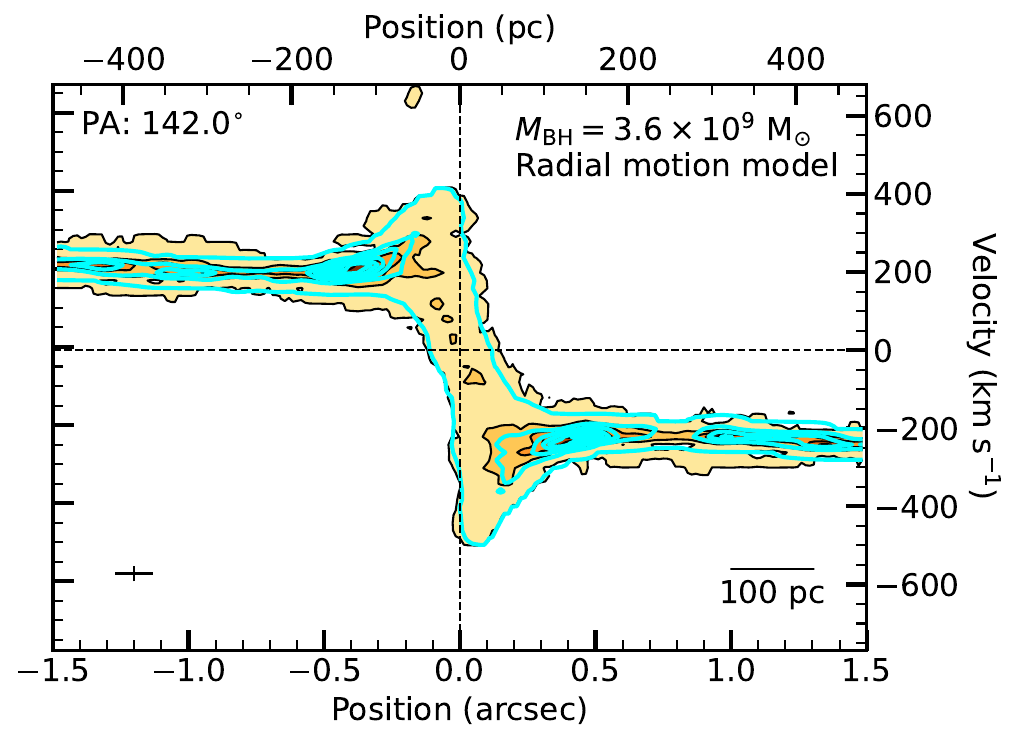}
    }
    \caption{Same as Figure~\ref{fig:natural_minor_pvd}, but showing the kinematic major-axis PVD of NGC~383 imaged with natural weighting. Both the warp and the radial motion model reproduce the forbidden quadrant emission slightly better than the model with no warp nor radial motion (in which all forbidden quadrant emission is due to beam smearing).
    }
    \label{fig:natural_major_pvd}
\end{figure*}

\subsubsection{Implications for the SMBH mass} \label{subsec:implications}

As neither a PA warp nor pure radial motion satisfactorily explains the observed velocity twist, that feature is more likely associated with more complicated non-circular motions (not necessarily pure radial inflow or outflow) that depend on the azimuthal angle and involve rotation velocities that are increased or decreased compared to those expected from pure circular motions. For example, the strong kink along the minor axis of the observed first-moment map, located at approximately ($-0\farcs1$, $0\farcs2$) from the centre, is reminiscent of those due to streaming motions along spiral arms or barred orbits. To search for such structures in the nuclear region, the central $1\farcs5\times1\farcs5$ of the zeroth-moment map of the data cube imaged with natural weighting is shown in the left panel of Figure~\ref{fig:zoomed_in_moment0}. The zeroth-moment map of an axisymmetric model generated using $^{\textsc{3D}}\textsc{Barolo}$ and the associated residuals ($\mathrm{data}-\mathrm{model}$) are also shown in the middle and right panels of Figure~\ref{fig:zoomed_in_moment0}, respectively. This reveals two arm-like structures of positive residuals near ($0\farcs25$, $-0\farcs5$) and ($0\farcs2$, $0\farcs45$), that may be evidence of a weak spiral structure. However, that structure is speculative and statistically insignificant compared to the substantial random and structured noise near the centre of the zeroth-moment map. Future observations with higher sensitivity are required to confirm the nature of this feature.
The large-scale spiral features observed by \citeauthor{North_2019} (\citeyear{North_2019}; see also the left panel of Figure~\ref{fig:moments}) are also recovered in the velocity residual map covering the entire molecular gas disc (Figure~\ref{fig:natural_velo_resi_full}). 
The tentative spiral-like structure in the zoomed-in zeroth-moment map is potentially a continuation of that large-scale spiral structure and could explain the velocity twist in the nuclear region.
Unfortunately, the relatively low $S/N$ and the limited number of pixels in the region of the velocity twist make it difficult to fit more complicated non-circular motion models to the data. In any case, we note that our best-fitting SMBH mass is not sensitive to potential non-circular motions, because of the dominance of the SMBH on the kinematics in the central region. All the models discussed above yield almost identical SMBH masses $M_\mathrm{BH}\approx3.6\times10^9$~M$_\odot$, and the best-fitting $M_\mathrm{BH}$ remains unchanged even if we mask out the twist region during the fit. Hence, we continue to adopt the results of the regular disc model in the following discussions, as it has the smallest BIC (Table~\ref{tab:model_stats}).

\begin{figure*}
\captionsetup[subfigure]{labelformat=empty}
    \centering
    \subfloat[]{
    \includegraphics[width=0.32\linewidth]{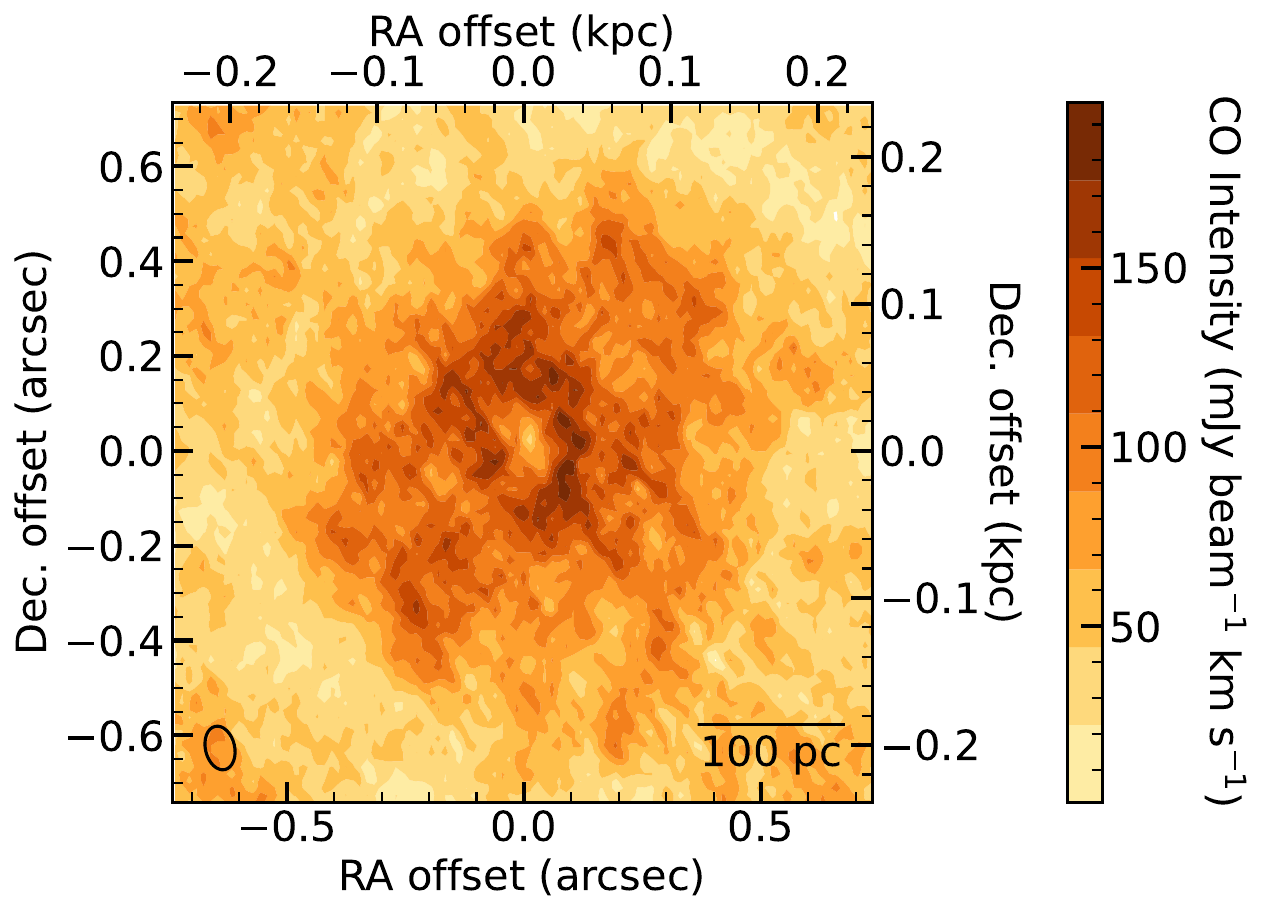}
    }
    \subfloat[]{
    \includegraphics[width=0.32\linewidth]{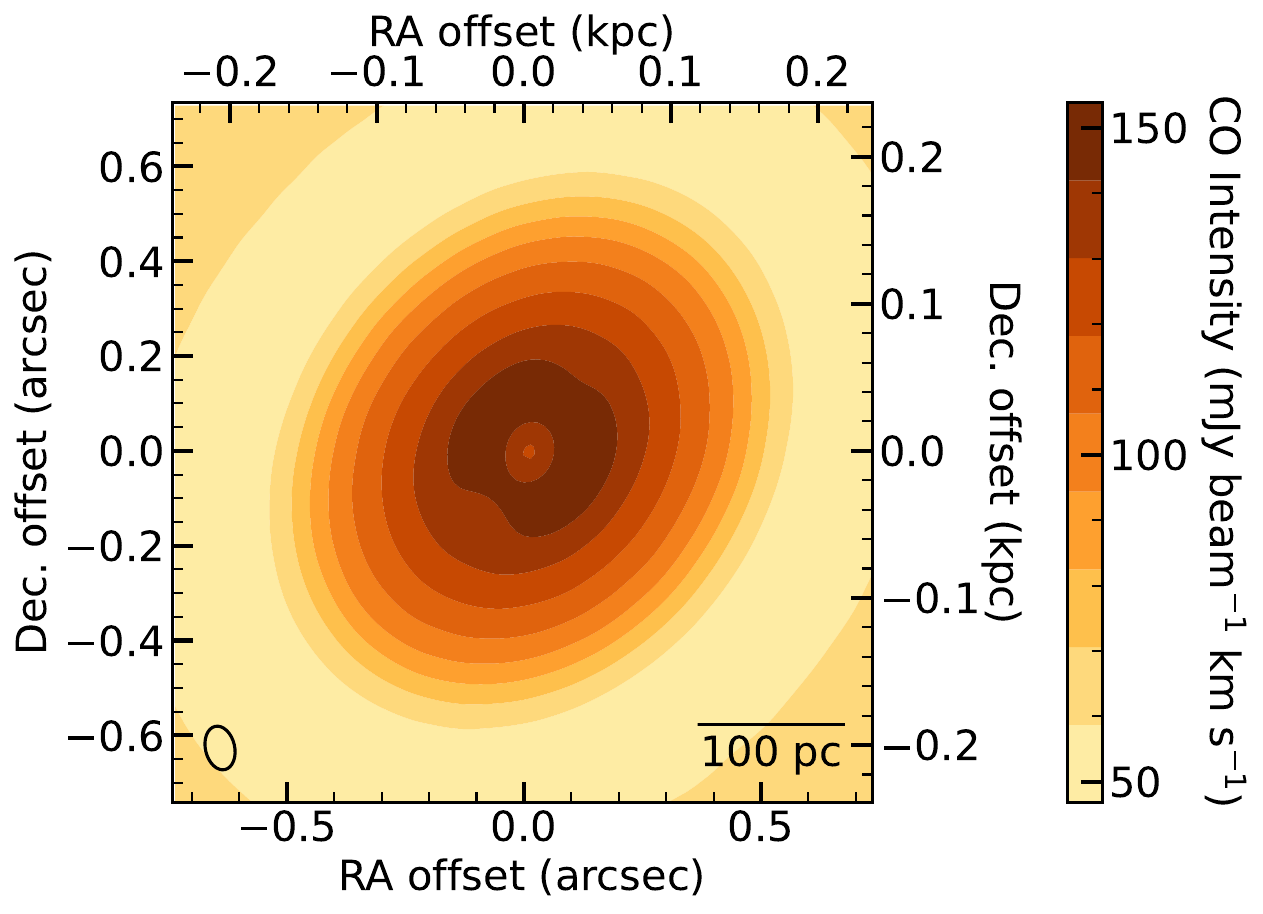}
    }
    \subfloat[]{
    \includegraphics[width=0.32\linewidth]{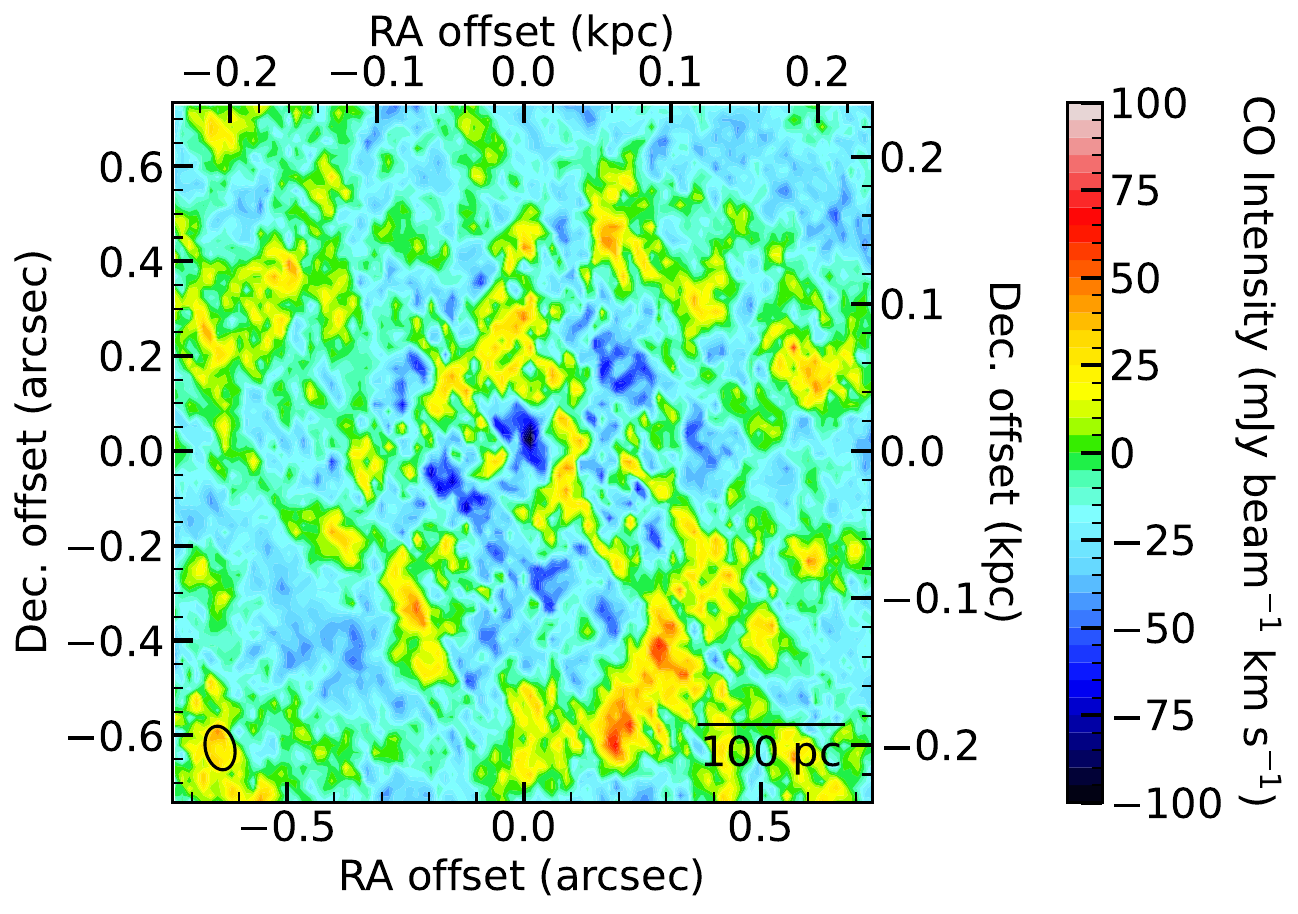}
    }
    \caption{Left: zeroth-moment (integrated-intensity) map of the NGC 383 data cube imaged with natural weighting, showing only the central $1\farcs5\times1\farcs5$. The dip within a radius of $\approx 0\farcs1$ is a physical decrease of the CO surface brightness. Middle: zeroth-moment map of the best-fitting axisymmetric model generated with
    $^{\textsc{3D}}\textsc{Barolo}$. The innermost few contours appear slightly asymmetric due to the elongation of the synthesised beam. Right: corresponding zeroth-moment residual map ($\mathrm{data}-\mathrm{model}$). The two arm-like structures of positive residuals near ($0\farcs25$, $-0\farcs5$) and ($0\farcs2$, $0\farcs45$) may be evidence of a weak spiral structure.}
    \label{fig:zoomed_in_moment0}
\end{figure*}

\begin{figure}
    \centering
    \includegraphics[width=\linewidth]{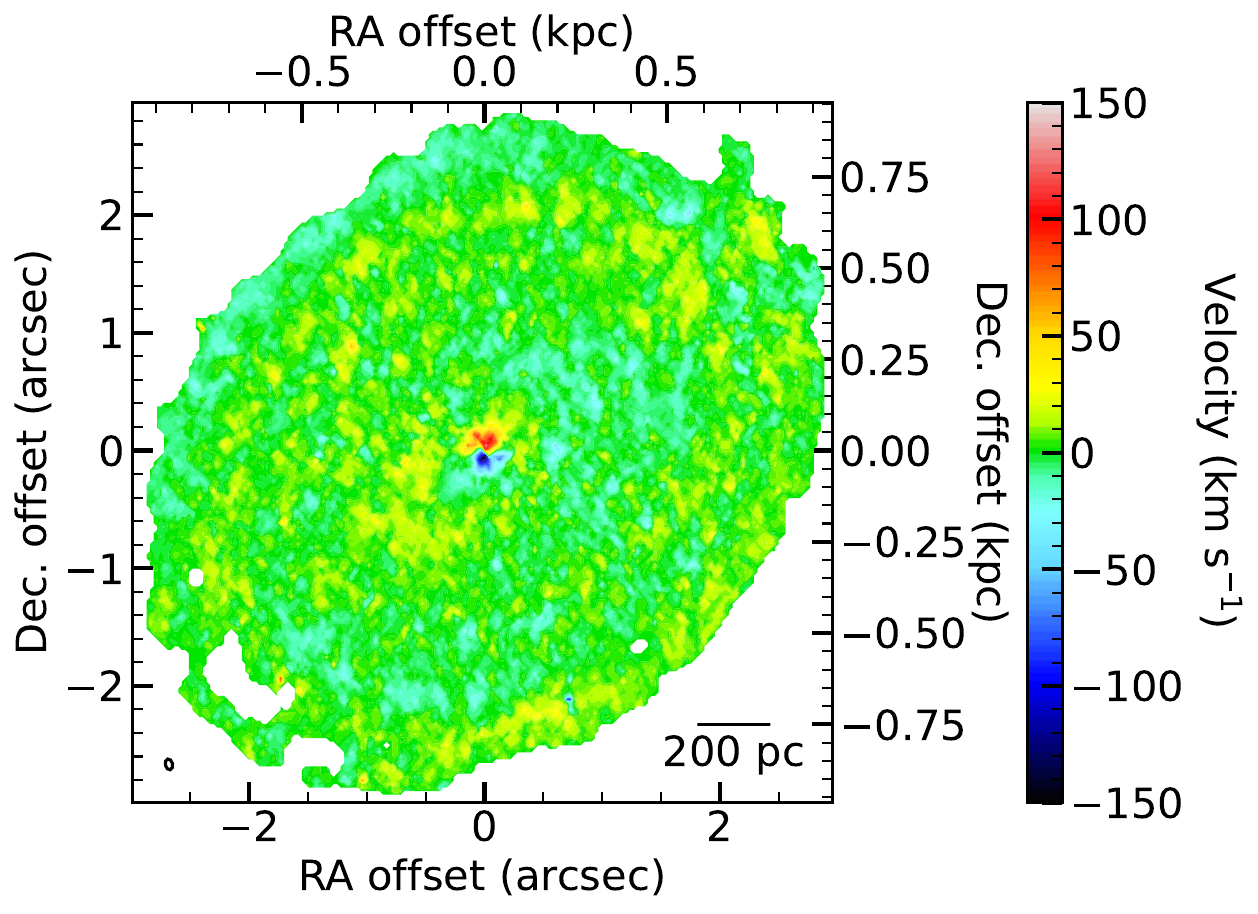}
    \caption{Full (i.e.\ entire disc) first-moment (i.e.\ intensity-weighted mean line-of-sight velocity) residual map of NGC~383, created by subtracting the first-moment map of the best-fitting model with no warp nor radial motion from the first-moment map of the data cube imaged with natural weighting. Apart from the large residuals at the centre, the large-scale spiral features observed by \citeauthor{North_2019} (\citeyear{North_2019}; see also the left panel of Figure~\ref{fig:moments}) are also recovered. This suggests non-circular motions more complicated than pure radial motions.}
    \label{fig:natural_velo_resi_full}
\end{figure}

Despite the evidence of non-circular motions, the asymmetry between the redshifted and blueshifted central velocity peaks of NGC~383 is unlikely to be caused by non-circular motions, as this feature is well reproduced by our best-fitting regular disc model assuming pure circular motions (see Figure~\ref{fig:pvd+model}). This implies that the asymmetry is most likely due to a deficiency of $^{12}$CO(2-1) emission on the redshifted side of the innermost region, that has already been incorporated into our model by construction as we use the observed (but deconvolved) intensity distribution as the input gas distribution. This deficit can have multiple origins, in particular, a non-axisymmetric gas morphology (e.g.\ the tentative nuclear spiral discussed above).


\subsection{Uncertainties} \label{subsec:uncertainty}

The typical uncertainties associated with a SMBH mass measurement using molecular gas kinematics have been discussed extensively in previous WISDOM papers \citep{Onishi_2017, Davis_2017, Davis_2018, Smith_2019, North_2019, Smith_2021, Lelli_2022, Ruffa_2023}. In the following, we thus discuss only the dominant sources of uncertainties of the SMBH mass of NGC~383, as derived from our high-resolution observations.


\subsubsection{$\chi^2$ rescaling}

As discussed in Section~\ref{sec:modelling}, we rescaled the uncertainties of the data cube by $(2N)^{0.25}$ to attempt to account for the potential systematic uncertainties usually dominating large data sets. Without $\chi^2$ rescaling, the SMBH mass precision would be $0.05\%$, an unrealistic uncertainty that severely underestimates potential systematic effects. To verify that this procedure provides a reasonable estimate of the systematic uncertainties and to confirm that the high-resolution data truly yields a more precise 
SMBH mass, we perform an independent estimate of the systematic uncertainties by adopting a procedure analogous to bootstrapping. We divide the data cube into four sub-samples, each one a spatial quadrant of the data cube, bounded by planes of fixed position angles. We fit each sub-sample independently, keeping the kinematic centre position and the systemic velocity fixed to their best-fitting values in Table~\ref{tab:results}. For each model parameter, we then adopt the mean of the four best-fitting parameters as the overall best fit, and their standard deviation as the $1\sigma$ uncertainty.
Table~\ref{tab:bootstrapp} compares the best-fitting parameters and the associated $1\sigma$ uncertainties obtained using this approach to those obtained in Section \ref{subsec:results} using the MCMC approach with $\chi^2$ rescaling. The best-fitting parameters obtained using the bootstrapping-like approach all agree within $1\sigma$ with those obtained using the MCMC approach and the uncertainties are also very similar. We thus confirm that the $\chi^2$ rescaling adopted provides reasonable estimates of the systematic uncertainties of all (non-nuisance) parameters of our model.

\begin{table}
    \centering
    \caption{Best-fitting parameters and associated $1\sigma$ uncertainties, obtained via MCMC with $\chi^2$ rescaling and bootstrapping, respectively.}
    \begin{tabular}{lcc} \hline
        Parameter & $\chi^2$ rescaling & Bootstrapping \\ \hline
        \multicolumn{3}{l}{\textbf{Mass model}} \\ $\log\left(M_\mathrm{BH}/\mathrm{M}_\odot\right)$ & \changes{$9.554 \pm 0.022$} & \changes{$9.571 \pm 0.024$} \\
        Inner stellar $M/L$ ($\mathrm{M}_\odot/\mathrm{L}_{\odot\mathrm{,F160W}}$) & \changes{$3.16 \pm 0.15$} & \changes{$3.04 \pm 0.17$} \\
        Outer stellar $M/L$ ($\mathrm{M}_\odot/\mathrm{L}_{\odot\mathrm{,F160W}}$) & \changes{$2.32 \pm 0.09$} & \changes{$2.32 \pm 0.10$} \\ \hline
        \multicolumn{3}{l}{\textbf{Molecular gas disc}} \\ 
        Position angle (degree) & $142.01 \pm 0.29$ & \changes{$142.19 \pm 0.27$} \\
        Inclination (degree) & $37.6 \pm 1.0$ & \changes{$38.0 \pm 1.0$} \\
        Velocity dispersion (\kms) & \changes{$10.6 \pm 0.9$} & \changes{$10.2 \pm 0.9$} \\ \hline
    \end{tabular}
    \label{tab:bootstrapp}
\end{table}


\subsubsection{Other uncertainties}
\label{subsec:other_uncertainty}

As our observations spatially resolve the SMBH's SoI by a factor of $\approx24$ radially (estimated using our updated SMBH mass and $\sigma_\mathrm{e}$, but see Section~\ref{subsec:high_res_merit} for a more physically-motivated calculation), they probe the innermost region of the molecular gas disc where the kinematics is dominated by the SMBH and is essentially independent of the stellar mass distribution. Consequently, the inferred SMBH mass depends only weakly on the stellar $M/L$. The positive correlation between SMBH mass and outer $M/L$ (see Figure~\ref{fig:corner}) is likely a by-product of the SMBH mass -- inclination (anti-)correlation and the $M/L$ -- inclination (anti-)correlation. We verified that a model with a spatially-constant $M/L$ does not improve the fit quality (using the BIC as the metric) and yields a SMBH mass $\log\left(M_\mathrm{BH}/\mathrm{M}_\odot\right)=9.58\pm0.02$ consistent with that from our radially linearly-varying $M/L$ model.

The SMBH mass of NGC~383 is however strongly degenerate with the inclination of the molecular gas disc. As $M_\mathrm{BH}\propto\sin^{-2}i$, the SMBH mass uncertainty is often dominated by the inclination uncertainty, especially when the gas disc is nearly face-on. With an inclination of only $38\degree$, the inclination uncertainty contributes $\approx55\%$ to the NGC~383 SMBH mass uncertainty budget. If the inclination were fixed to the best-fitting inclination, the SMBH mass precision would be $\approx4\%$ for the high-resolution data cube fit and $\approx9\%$ for the intermediate-resolution data cube fit.

We note that every uncertainty discussed so far is much smaller than that introduced by the $\approx15\%$ uncertainty of the adopted distance ($66.6\pm9.9$~Mpc; \citealt{Freedman_2001}). Indeed, as the SMBH mass scales linearly with the adopted distance, the distance uncertainty can impact the SMBH mass more than all other uncertainties combined. Nevertheless, as scaling the SMBH mass to a different distance is straightforward (i.e.\ it does not require redoing the fit), we follow the customary practice of not including the distance uncertainty in our results.


\subsection{Merits of high-resolution observations}
\label{subsec:high_res_merit}

Our high-resolution observations of the molecular gas disc of NGC~383 spatially resolve material with a maximum line-of-sight velocity $V_\mathrm{obs}\approx635$~\kms, higher than the $\approx350$~\kms\ detected by the intermediate-resolution observations of \citet{North_2019}. Deprojecting this with $V_\mathrm{c}=V_\mathrm{obs}/\sin{i}$, where the best-fitting inclination $i=37\fdg6$, the highest-velocity material resolved by our observations has a circular velocity $V_\mathrm{c}\approx1040$~\kms. This is larger than the maximum circular velocities probed by all previous molecular gas kinematic SMBH mass measurements \citep{Zhang_2024}. This largest $V_\mathrm{c}$ is the result of our measurement resolving material physically closest to the SMBH in terms of the number of Schwarzschild radii ($R/R_\mathrm{Schw}\approx4.1\times10^4$), as $\left(V_\mathrm{c}/c\right)=\sqrt{2}\left(R_\mathrm{Schw}/R\right)^{1/2}$ for a Keplerian circular velocity curve \citep{North_2019}. Our measurement also has the largest $R_\mathrm{SoI}/R$ ratio ($R_\mathrm{SoI}/R\approx24$; \citealt{Zhang_2024}), thus spatially resolving the SoI (defined using the effective stellar velocity dispersion $\sigma_\mathrm{e}$) better than all existing molecular gas SMBH mass measurements.

We also consider an alternative definition of the SoI radius, the radius $R_\mathrm{eq}$ at which the enclosed stellar mass equals the SMBH mass:

\begin{equation}
    M_*(R=R_\mathrm{eq})=M_\mathrm{BH}\,\,.
\end{equation}
This $R_\mathrm{eq}$ is a more accurate definition of the SoI radius than $R_\mathrm{SoI}\equiv GM_\mathrm{BH}/\sigma_\mathrm{e}^2$, as $\sigma_\mathrm{e}$ is only a proxy of the galaxy's gravitational potential and thus the stellar mass at large spatial scales. By contrast, $R_\mathrm{eq}$ is defined from an explicit comparison of the SMBH mass and the stellar mass. Using the MGE model in Table~\ref{tab:MGE} and the best-fitting $M/L$ and SMBH mass, we derive $R_\mathrm{eq}=\changes{0\farcs88}\pm0\farcs02$ ($\changes{285}\pm6$~pc), in good agreement with \changes{$R_\mathrm{SoI} = 0\farcs84 \pm 0.05$}. Our observations thus resolve $R_\mathrm{eq}$ by a factor of \changes{$\approx25$}, again better than all prior molecular gas SMBH measurements \citep{Zhang_2024}.

In other words, using either $R_\mathrm{Schw}$, $R_\mathrm{SoI}$ or $R_\mathrm{eq}$ as the reference spatial scale, our measurement is the highest-resolution SMBH mass measurement using molecular gas kinematics to date. We note that our observations spatially resolve material closer to the SMBH in the unit of $R_\mathrm{Schw}$ than even the best SMBH mass measurement using masers ($R/R_\mathrm{Schw}\approx4.4\times10^4$ for NGC~4258; \citealt{Herrnstein_2005}).

The high spatial resolution of our observations results in a SMBH mass precision of $5\%$ after $\chi^2$ rescaling, consistent with the uncertainty estimated via bootstrapping. 
The SMBH mass uncertainty can not be directly compared to those of previous SMBH mass measurements, as each measurement adopts a different approach to incorporate systematic uncertainties. Among the measurements that use $\chi^2$ rescaling and do not fix the inclination, our SMBH mass precision is higher than all but those of NGC~3557 ($5\%$; \citealt{Ruffa_2019b}) and NGC~7052 ($4\%$; \citealt{Smith_2021}). This demonstrates again the importance of high spatial resolution for a precise SMBH mass. 

NGC~383's SMBH mass uncertainty arises primarily from the inclination uncertainty. The mass precision would improve more substantially with spatial resolution for a less face-on disc. We note that the precision of a SMBH measurement also depends on the $S/N$, but at a given $S/N$ the ALMA integration time ($t_\mathrm{int}$) is inversely proportional to the fourth power of the synthesised beam's FWHM ($\theta_\mathrm{beam}$): $t_\mathrm{int}\propto\theta_\mathrm{beam}^{-4}$. Therefore, ultra-high-resolution observations require much longer integration times that limit their schedulability, or one must make do with lower $S/N$. With the current high-resolution observations, the SMBH mass precision is impaired by the limited $S/N$. If our high-resolution observations had achieved the same $S/N$ as the existing intermediate-resolution observations, the mass precision would have been $\approx4\%$, estimated by re-running the MCMC assuming a lower $\sigma_\mathrm{RMS}$. To summarise, higher spatial resolution observations should lead to better SMBH mass precision at a given inclination and $S/N$, but achieving the desired $S/N$ is more costly.

A higher spatial resolution should improve not only the precision but also the accuracy of a SMBH mass measurement, as it can reveal features that can bias the kinematic modelling but often remain unresolved and thus undetected at lower resolutions. For example, our high-resolution observations of NGC~383 reveal an asymmetry of the central blueshifted and redshifted velocity peaks and an offset between the kinematic and morphological centres, both unseen in the intermediate-resolution observations. By incorporating both features in our model, we reduced the inaccuracy of our kinematic modelling. The best-fitting SMBH mass changed slightly from $(4.2\pm0.4)\times10^9$ to $(3.6\pm0.2)\times10^9$~M$_\odot$ (both $1\sigma$ uncertainties), although the two masses are statistically consistent. The high-resolution observations also unveiled a velocity twist within the central $\approx0\farcs3$ in radius, thus providing evidence of non-circular motions in the circumnuclear disc. Although the best-fitting $M_\mathrm{BH}$ is unaffected in this case because the twist is limited to the central region where the SMBH dominates the kinematics, an accurate model of non-circular motions could substantially improve the SMBH mass constraint for similar observations of other galaxies \citep[e.g.][]{Lelli_2022}. 

\changes{We however caution that a higher spatial resolution does not guarantee a more precise and accurate SMBH mass measurement when there is a central dip or hole in the CO surface brightness distribution. Although the slight central dip of NGC~383 does not affect our measurement, a central hole in the CO distribution often prevents high-resolution observations from probing emission sufficiently close to the SMBH and thus prohibits a precise SMBH mass measurement \citep[e.g.][]{Kabasares_2022, Ruffa_2023}. Nevertheless, one may be able to adopt a different kinematic tracer, such as [CI], which often extends closer to the SMBH than CO and does not have a central hole in its distribution \citep[e.g.][]{Izumi_2020, Nguyen_2021}, to obtain a precise SMBH mass measurement even in the presence of a central CO hole.}


\section{Conclusions}
\label{sec:conclusion}

We presented a measurement of the SMBH mass of NGC~383, a lenticular galaxy hosting a jetted AGN. We used ALMA observations of the $^{12}$CO(2-1) emission line with a synthesised beam FWHM of $0\farcs050\times0\farcs024$ ($\approx16\times8$~pc$^2$), $\approx4$ times better than previous observations, spatially resolving the SMBH SoI by a factor of $\approx24$ in radius. This spatial resolution, in the unit of either $R_\mathrm{Schw}$, $R_\mathrm{SoI}$ or $R_\mathrm{eq}$, is the highest of all SMBH mass measurements using molecular gas. The observations also spatially resolve material closer to the SMBH in terms of the number of Schwarzschild radii \changes{($R/R_\mathrm{Schw}\approx4.1\times10^4$)} than the best measurements using masers \changes{($R/R_\mathrm{Schw}\approx4.4\times10^4$ for NGC~4258; \citealt{Herrnstein_2005})}. Our high-resolution data detect a clear Keplerian rise (from the outside in) of the (deprojected) circular velocities up to $\approx1040$~\kms. They also reveal features undetected in previous observations: an offset between the gas disc's kinematic and morphological centres, a mild asymmetry in the innermost velocity peaks and a central velocity twist suggesting non-circular motions. Our best-fitting kinematic model yields a SMBH mass of \changes{$(3.58\pm0.19)\times10^9$}~M$_\odot$, more precise but consistent within $\approx1.4\sigma$ with the previous SMBH mass of $(4.2\pm0.4)\times10^9$~M$_\odot$ derived from intermediate-resolution data \citep{North_2019}. This SMBH mass is insensitive to models of the central velocity twist. The F160W-filter stellar $M/L$ decreases from \changes{$3.16\pm0.15$}~$\mathrm{M}_\odot/\mathrm{L}_{\odot\mathrm{,F160W}}$ at the centre to \changes{$2.32\pm0.09$}~$\mathrm{M}_\odot/\mathrm{L}_{\odot\mathrm{,F160W}}$ at the outer edge of the molecular gas disc ($R=3\farcs5$).

Despite this highest spatial resolution, this SMBH mass measurement is not the most precise because of the relatively low inclination, associated with a high inclination uncertainty, and the relatively low $S/N$. It nevertheless demonstrates the importance of high resolution to the precision and accuracy of SMBH mass measurements. Although high-resolution observations require relatively long ALMA integration times, they will steadily increase the number of precise and accurate SMBH mass measurements using molecular gas kinematics, and they will ultimately rival the "gold standard" measurements using masers. This will, in turn, allow constraining SMBH -- galaxy scaling relations more tightly across the entire galaxy range.


\section*{Acknowledgements}

HZ acknowledges support from a Science and Technology Facilities Council (STFC) DPhil studentship under grant ST/X508664/1 and the Balliol College J.\ T.\ Hamilton Scholarship in physics.
MB was supported by STFC consolidated grant ‘Astrophysics at Oxford’ ST/K00106X/1 and ST/W000903/1. IR and TAD acknowledge support from STFC consolidated grant ST/S00033X/1.
This paper makes use of the following ALMA data: ADS/JAO.ALMA\#2015.1.00419.S, ADS/JAO.ALMA\#2016.1.00437.S and ADS/JAO.ALMA\#2019.1.00582.S. ALMA is a partnership of ESO (representing its member states), NSF (USA) and NINS (Japan), together with NRC (Canada), MOST and ASIAA (Taiwan), and KASI (Republic of Korea), in cooperation with the Republic of Chile. The Joint ALMA Observatory is operated by ESO, AUI/NRAO and NAOJ. This research made use of the NASA/IPAC Extragalactic Database (NED), which is operated by the Jet Propulsion Laboratory, California Institute of Technology, under contract with the National Aeronautics and Space Administration.


\section*{Data Availability}

The raw data used in this study are publicly available on the ALMA Science Archive: \url{https://almascience.eso.org/aq/}, and on the Hubble Science Archive: \url{https://hst.esac.esa.int/ehst/}. The calibrated data, final data products and original plots generated for this research will be shared upon reasonable requests to the
first author.


\bibliographystyle{mnras}
\bibliography{refs} 

\bsp	
\label{lastpage}
\end{document}